\PassOptionsToPackage{unicode}{hyperref}
\PassOptionsToPackage{hyphens}{url}
\documentclass[
  10pt,
]{article}
\usepackage{amsmath,amssymb}
\usepackage{lmodern}
\usepackage{setspace}
\usepackage{ifxetex,ifluatex}
\ifnum 0\ifxetex 1\fi\ifluatex 1\fi=0 
  \usepackage[T1]{fontenc}
  \usepackage[utf8]{inputenc}
  \usepackage{textcomp} 
\else 
  \usepackage{unicode-math}
  \defaultfontfeatures{Scale=MatchLowercase}
  \defaultfontfeatures[\rmfamily]{Ligatures=TeX,Scale=1}
\fi
\IfFileExists{upquote.sty}{\usepackage{upquote}}{}
\IfFileExists{microtype.sty}{
  \usepackage[]{microtype}
  \UseMicrotypeSet[protrusion]{basicmath} 
}{}
\makeatletter
\@ifundefined{KOMAClassName}{
  \IfFileExists{parskip.sty}{%
    \usepackage{parskip}
  }{
    \setlength{\parindent}{0pt}
    \setlength{\parskip}{6pt plus 2pt minus 1pt}}
}{
  \KOMAoptions{parskip=half}}
\makeatother
\usepackage{xcolor}
\IfFileExists{xurl.sty}{\usepackage{xurl}}{} 
\IfFileExists{bookmark.sty}{\usepackage{bookmark}}{\usepackage{hyperref}}
\hypersetup{
  pdftitle={An Efficient Method for Computing Expected Value of Sample Information for Survival Data from an Ongoing Trial},
  hidelinks,
  pdfcreator={LaTeX via pandoc}}
\usepackage[margin=25mm]{geometry}
\usepackage{longtable,booktabs,array}
\usepackage{calc} 
\usepackage{etoolbox}
\makeatletter
\patchcmd\longtable{\par}{\if@noskipsec\mbox{}\fi\par}{}{}
\makeatother
\IfFileExists{footnotehyper.sty}{\usepackage{footnotehyper}}{\usepackage{footnote}}
\makesavenoteenv{longtable}
\usepackage{graphicx}
\makeatletter
\def\maxwidth{\ifdim\Gin@nat@width>\linewidth\linewidth\else\Gin@nat@width\fi}
\def\maxheight{\ifdim\Gin@nat@height>\textheight\textheight\else\Gin@nat@height\fi}
\makeatother
\setkeys{Gin}{width=\maxwidth,height=\maxheight,keepaspectratio}
\makeatletter
\def\fps@figure{htbp}
\makeatother
  \usepackage{float}  
  \usepackage{booktabs}
  \usepackage{bm}
  \usepackage{xcolor}
  \usepackage{nicefrac}
  \usepackage{mathtools}
  \usepackage[T1]{fontenc}
  \usepackage[colorinlistoftodos]{todonotes} 
  \usepackage{soul} 

  \usepackage[ruled,vlined]{algorithm2e} 

  \newcommand{\EVSI}{\mbox{$\operatorname{EVSI}$}}
  \newcommand{\NB}{\mbox{$\mathrm{NB}$}}
  \newcommand{\E}{\mbox{$\mathbb{E}$}}

\newcommand{\beginappendixA}{%
        \setcounter{table}{0}
        \renewcommand{\thetable}{A\arabic{table}}%
        \setcounter{figure}{0}
        \renewcommand{\thefigure}{A\arabic{figure}}%
     }

\newcommand{\beginappendixB}{%
        \setcounter{table}{0}
        \renewcommand{\thetable}{B\arabic{table}}%
        \setcounter{figure}{0}
        \renewcommand{\thefigure}{B\arabic{figure}}%
     }
     
\newcommand{\beginappendixC}{%
        \setcounter{table}{0}
        \renewcommand{\thetable}{C\arabic{table}}%
        \setcounter{figure}{0}
        \renewcommand{\thefigure}{C\arabic{figure}}%
     }     
     
\newcommand{\beginappendixD}{%
        \setcounter{table}{0}
        \renewcommand{\thetable}{D\arabic{table}}%
        \setcounter{figure}{0}
        \renewcommand{\thefigure}{D\arabic{figure}}%
     }     

\newcommand{\beginappendixE}{%
        \setcounter{table}{0}
        \renewcommand{\thetable}{E\arabic{table}}%
        \setcounter{figure}{0}
        \renewcommand{\thefigure}{E\arabic{figure}}%
     }

\newcommand{\beginappendixF}{%
        \setcounter{table}{0}
        \renewcommand{\thetable}{F\arabic{table}}%
        \setcounter{figure}{0}
        \renewcommand{\thefigure}{F\arabic{figure}}%
     } 
     
\newcommand{\beginappendixG}{%
        \setcounter{table}{0}
        \renewcommand{\thetable}{G\arabic{table}}%
        \setcounter{figure}{0}
        \renewcommand{\thefigure}{G\arabic{figure}}%
     }

\makeatletter
\g@addto@macro{\table}{\centering}
\makeatother
\usepackage{booktabs}
\usepackage{longtable}
\usepackage{array}
\usepackage{multirow}
\usepackage{wrapfig}
\usepackage{float}
\usepackage{colortbl}
\usepackage{pdflscape}
\usepackage{tabu}
\usepackage{threeparttable}
\usepackage{threeparttablex}
\usepackage[normalem]{ulem}
\usepackage{makecell}
\usepackage{xcolor}
\ifluatex
  \usepackage{selnolig}  
\fi
\newlength{\cslhangindent}
\newlength{\csllabelwidth}
\newenvironment{CSLReferences}[2] 
 {
  \setlength{\parindent}{0pt}
  \ifodd #1 \everypar{\setlength{\hangindent}{\cslhangindent}}\ignorespaces\fi
  \ifnum #2 > 0
  \setlength{\parskip}{#2\baselineskip}
  \fi
 }%
 {}
\usepackage{calc}

\newcommand{\CSLLeftMargin}[1]{\parbox[t]{\csllabelwidth}{#1}}
\newcommand{\CSLRightInline}[1]{\parbox[t]{\linewidth - \csllabelwidth}{#1}\break}

\title{An Efficient Method for Computing Expected Value of Sample Information for Survival Data from an Ongoing Trial}
\author{Mathyn Vervaart, MSc, MPhil\(^{1,2,}\)\footnote{\textbf{Corresponding author:} Mathyn Vervaart, Department of Health Management and Health Economics, University of Oslo, Forskningsveien 3A, Harald Schjelderups hus, 0373 Oslo, Norway (\href{mailto:mathyn.vervaart@medisin.uio.no}{\nolinkurl{mathyn.vervaart@medisin.uio.no}}).} , Mark Strong, PhD\(^{3}\), Karl P. Claxton, PhD\(^{4,5}\), \\
Nicky J. Welton, PhD\(^{6}\), Torbjørn Wisløff, PhD\(^{7,8}\), Eline Aas, PhD\(^{1}\)\\
~\\
\small \(^1\)Department of Health Management and Health Economics, University of Oslo, Oslo, Norway\\
\small \(^2\)Norwegian Medicines Agency, Oslo, Norway\\
\small \(^3\)School of Health and Related Research, University of Sheffield, Sheffield, UK\\
\small \(^4\)Centre for Health Economics, University of York, York, UK\\
\small \(^5\)Department of Economics and Related Studies, University of York, York, UK\\
\small \(^6\)Population Health Sciences, University of Bristol, Bristol, UK\\
\small \(^7\)Department of Community Medicine, UiT The Arctic University of Norway, Oslo, Norway\\
\small \(^8\)Norwegian Institute of Public Health, Oslo, Norway}
\date{\small 25 March, 2021}

\begin{document}
\maketitle
\begin{abstract}
\noindent The European Medicines Agency has in recent years allowed licensing of new pharmaceuticals at an earlier stage in the clinical trial process.
When trial evidence is obtained at an early stage, the events of interest, such as disease progression or death, may have only been observed in a small proportion of patients.
Health care authorities therefore must decide on the adoption of new technologies based on less mature evidence than previously, resulting in greater uncertainty about clinical- and cost-effectiveness.
When a trial is ongoing at the point of decision making, there may be value in continuing the trial in order to collect additional data before making an adoption decision. This can be quantified by the Expected Value of Sample Information (EVSI).
However, no guidance exists on how to compute the EVSI for survival data from an ongoing trial, nor on how to account for uncertainty about the choice of survival model in the EVSI calculations.
In this article, we describe algorithms for computing the EVSI of extending a trial's follow-up, both where a single known survival model is assumed, and where we are uncertain about the true survival model.
We compare a nested Markov Chain Monte Carlo procedure with a non-parametric regression-based method in two synthetic case studies, and find close agreement between the two methods.
The regression-based method is fast and straightforward to implement, and scales easily to include any number of candidate survival models in the model uncertainty case.
EVSI for ongoing trials can help decision makers determine whether early patient access to a new technology can be justified on the basis of the current evidence or whether more mature evidence is needed.
\end{abstract}

\setstretch{1}
\hypertarget{intro}{%
\section{Introduction}\label{intro}}

The Expected Value of Sample Information (EVSI) quantifies the expected value to the decision maker of reducing uncertainty through the collection of additional data,\textsuperscript{\protect\hyperlink{ref-schlaiferProbabilityStatisticsBusiness1959}{1},\protect\hyperlink{ref-raiffaAppliedStatisticalDecision1961a}{2}} for example a \textit{future} randomised controlled trial.
Although a few studies have considered the use of EVSI methods at interim analyses of adaptive trials,\textsuperscript{\protect\hyperlink{ref-flightReviewClinicalTrials2019}{3}} overall little research has been done on EVSI for trials that are ongoing at the point of decision making.

In the last decade the European Medicines Agency (EMA) has introduced regulatory mechanisms that are aimed at accelerating licensing of new pharmaceuticals, such as `adaptive pathways'\textsuperscript{\protect\hyperlink{ref-europeanmedicinesagencyAdaptivePathways2018}{4}} and `conditional marketing authorisations.'\textsuperscript{\protect\hyperlink{ref-europeanmedicinesagencyConditionalMarketingAuthorisation2018}{5}}
When evidence is obtained from a trial at an early stage, the events of interest, such as disease progression or death, may have only been observed in a small proportion of patients.
Health care authorities therefore have to issue guidance on new pharmaceuticals based on less mature evidence than previously, resulting in greater uncertainty about clinical- and cost-effectiveness.
With this comes an increased risk of recommending a technology that \textit{reduces} net health benefit.\textsuperscript{\protect\hyperlink{ref-claxtonComprehensiveAlgorithmApproval2016}{6}}

Additional evidence can be valuable as it can lead to better decisions that improve health and/or reduce resource use.\textsuperscript{\protect\hyperlink{ref-claxtonComprehensiveAlgorithmApproval2016}{6}}
Positive adoption decisions can be costly or difficult to reverse, and may remove the incentives for manufacturers to provide additional data.
When a trial is ongoing at the point of decision making, for example when follow-up is continued for regulatory purposes, there may therefore be value in delaying the adoption decision until additional data has been collected in the ongoing trial and uncertainty has reduced.\textsuperscript{\protect\hyperlink{ref-eckermannOptionValueDelay2008}{7}}
In this context, there will be a trade-off between granting early access to a new technology that may turn out to reduce health benefits, and waiting for uncertainty to be reduced through ongoing data collection with a potential loss of health benefits while waiting.
When the manufacturer is already committed to continuing the ongoing trial, the option to delay a decision is relevant even in a policy context where the decision maker does not have the formal authority to commission research.
The value of delaying the decision could be quantified, at least in theory, by computing the EVSI for the additional follow-up data.

Estimates of life-expectancy and time to disease progression are often key drivers of cost-effectiveness, particularly in oncology.
However, immature data means that there may be substantial uncertainty around these estimates and they rely on extrapolation beyond the trial follow-up period.\textsuperscript{\protect\hyperlink{ref-gallacherExtrapolatingParametricSurvival2020}{8}}
The choice of the survival distribution for extrapolation can have major implications for cost-effectiveness, and uncertainty surrounding this choice can be accounted for by model averaging, which may improve the quality of the extrapolations compared to selecting a single model.\textsuperscript{\protect\hyperlink{ref-gallacherExtrapolatingParametricSurvival2021}{9}}
A potential benefit of continuing an ongoing trial is to reduce the structural uncertainty as to the most appropriate survival distribution.
However, to the best of the authors knowledge, there exists no guidance on how to compute EVSI for survival data from a trial that is ongoing at the point of decision making, nor on how to account for structural uncertainty about the choice of survival model in the EVSI calculations.

In this article, we present algorithms for computing the EVSI of extending a trial's follow-up with and without accounting for structural uncertainty.
The algorithms are based on nested Markov Chain Monte Carlo methods and a fast nonparametric regression-based method.\textsuperscript{\protect\hyperlink{ref-strongEstimatingExpectedValue2015}{10}}
The nonparametric regression-based method\textsuperscript{\protect\hyperlink{ref-strongEstimatingExpectedValue2015}{10}} is generally more practical than other EVSI approximation methods as it neither requires nested Monte Carlo computations nor importance sampling.\textsuperscript{\protect\hyperlink{ref-heathCalculatingExpectedValue2020}{11}}
The article is structured as follows.
In the second section, we describe single-model and model-averaged EVSI algorithms for survival data from an ongoing trial.
In the third section, we compare the EVSI algorithms in two illustrative case studies, and in a final section, conclude with a brief discussion.

\hypertarget{method}{%
\section{Method}\label{method}}

\hypertarget{evsi-for-an-ongoing-study-collecting-time-to-event-data}{%
\subsection{EVSI for an ongoing study collecting time-to-event data}\label{evsi-for-an-ongoing-study-collecting-time-to-event-data}}

\hypertarget{decision-problem-and-model-definition}{%
\subsubsection{Decision problem and model definition}\label{decision-problem-and-model-definition}}

We assume a decision problem with \(d = 1,\ldots, D\) decision options.
The net benefit of option \(d\) is \(\NB(d,\bm\theta)\), and we have a cost-effectiveness model that predicts this quantity, given a vector of \(p\) possibly correlated model input parameters, \(\bm\theta = \{\theta_1,\ldots,\theta_p\}\).
Our current judgements about the vector \(\bm\theta\) is represented by the joint probability distribution \(p(\bm\theta)\).
Our goal is to choose the decision option with the greatest net benefit.

\hypertarget{evsi-for-further-follow-up-in-an-ongoing-study}{%
\subsubsection{EVSI for further follow-up in an ongoing study}\label{evsi-for-further-follow-up-in-an-ongoing-study}}

The EVSI for a new study that will provide (as yet uncollected) data, \(\mathbf{x}\), is defined as:
\begin{equation}
\EVSI\mathrm{(new\;study)}=\E_\mathbf{x}[\max_d\E_{\bm\theta|\mathbf{x}} \{ \NB(d, \bm\theta) \}] - \max_d\E_{\bm\theta} \{ \NB(d, \bm\theta) \},
\label{eq:evsi1}
\end{equation}
where the first term is the expected value of a decision based on our beliefs about \(\bm\theta\) given the new data, \(p(\bm\theta|\mathbf{x})\), and the second term is the expected value of a decision based on our beliefs about \(\bm\theta\) given current information alone, \(p(\bm\theta)\).\textsuperscript{\protect\hyperlink{ref-adesExpectedValueSample2004}{12}}
We now imagine that data \(\mathbf{x}\) have been collected during a given follow-up period for this study, which we denote time \(t_1\).
This could be an interim analysis, or the end of the study follow period.

The value of extending follow-up from current time \(t_1\) to some future point \(t_2\) is given by
\begin{equation}
\EVSI\mathrm{(ongoing\;study)}=\E_{\mathbf{\tilde{x}}|\mathbf{x}}[\max_d\E_{\bm\theta|\mathbf{x},\mathbf{\tilde{x}}} \{ \NB(d, \bm\theta) \}] - \max_d\E_{\bm\theta|\mathbf{x}} \{ \NB(d, \bm\theta) \},
\label{eq:evsi_ongoing}
\end{equation}
where the first term is the expected value of a decision based on our beliefs about \(\bm\theta\) given both new data, \(\mathbf{\tilde{x}}\), collected between \(t_1\) and \(t_2\), and data, \(\mathbf{x}\), collected between time zero and \(t_1\).
The second term is the expected value of a decision based on our beliefs about \(\bm\theta\) given only the information collected up until \(t_1\).
See Appendix A for a fuller explanation.

\hypertarget{specifying-current-beliefs-about-model-parameters-for-an-ongoing-study}{%
\subsubsection{Specifying current beliefs about model parameters for an ongoing study}\label{specifying-current-beliefs-about-model-parameters-for-an-ongoing-study}}

The distribution for the cost-effectiveness model parameters given knowledge at \(t_1\) \(p(\bm\theta|\mathbf{x})\) can be defined either in a fully Bayesian manner, by updating (possibly vague) prior information about \(\bm\theta\) with data \(\mathbf{x}\), or by fitting a standard frequentist statistical model to \(\mathbf{x}\) and obtaining the maximum likelihood estimate for \(\bm\theta\) along with some expression of uncertainty, and treating this as a Bayesian posterior.
In the absence of strong prior information about \(\bm\theta\), the two methods will produce very similar distributions for \(p(\bm\theta|\mathbf{x})\), even with relatively little data.\textsuperscript{\protect\hyperlink{ref-albertBayesianComputation2009}{13}}

\hypertarget{specifying-the-likelihood-for-ongoing-time-to-event-data-and-left-truncation}{%
\subsubsection{Specifying the likelihood for ongoing time-to-event data and left-truncation}\label{specifying-the-likelihood-for-ongoing-time-to-event-data-and-left-truncation}}

To compute EVSI we must define the data-generating distribution for the follow-up data between \(t_1\) and \(t_2\), \(p(\mathbf{\tilde{x}}|\bm\theta)\).
We first consider the structure of the data we will observe.
We assume our study has two arms: new treatment and standard care, and that \(N\) participants are recruited into each arm.
Data, \(\mathbf{x}\), collected from time zero to \(t_1\) take the form of a vector of times-to-death, -end of follow-up or -loss to follow-up, whichever is soonest.
Survival times for those alive at \(t_1\) are censored.
If we continue to collect data \(\mathbf{\tilde{x}}\) from \(t_1\) to \(t_2\) we may observe times-to-death for the participants whose observations were censored at \(t_1\).
Survival times for those alive at \(t_2\) or lost to follow-up are now the only observations censored.
Table \ref{tab:data-struct} illustrates the structure of the data for one arm of a study with follow-up at 12 and 24 months.

\begin{table}

\begin{threeparttable}
\caption{\label{tab:data-struct}The structure of the data for one arm of a study with follow-up at 12 and 24 months. Five participants are shown. Data are denoted \(\mathbf{x} = \{(9.3, 12, 12, 6.7, 12), (1, 0, 0, 0, 0)\}\) for observations up until \(t_1 = 12\) months, and \(\mathbf{\tilde{x}} =\{(13.4, 24, 15.9), (1, 0, 0)\}\) for observations between \(t_1\) and \(t_2 = 24\) months.}
\centering
\fontsize{7.8}{10}\selectfont
\begin{tabular}[t]{ccccccl}
\toprule
\multicolumn{1}{c}{ } & \multicolumn{3}{c}{Follow-up up at $t_1 = 12$ months} & \multicolumn{2}{c}{Follow-up at $t_2 = 24$ months} & \multicolumn{1}{c}{ } \\
\cmidrule(l{3pt}r{3pt}){2-4} \cmidrule(l{3pt}r{3pt}){5-6}
ID & Survival time & Censoring indicator, $\delta$ & At risk at $t_1$ & Survival time & Censoring indicator, $\delta$ & Outcome\\
\midrule
1 & 9.3 & 1 & No & - & - & Died at 9.3 months\\
2 & $12.0^*$ & 0 & Yes & 13.4 & 1 & Died at 13.4 months\\
3 & $12.0^*$ & 0 & Yes & $24.0^*$ & 0 & Alive at 24.0 months\\
4 & $6.7^*$ & 0 & No & - & - & LFU at 6.7 months\\
5 & $12.0^*$ & 0 & Yes & $15.9^*$ & 0 & LFU at 15.9 months\\
\bottomrule
\end{tabular}
\begin{tablenotes}
\small
\item [] *Observation censored $(\delta=0)$. LFU = Lost to follow-up.
\end{tablenotes}
\end{threeparttable}
\end{table}

Survival times are usually assumed to arise from a data generating process that can be described using a parametric model, the form of which must be chosen by the analyst.\textsuperscript{\protect\hyperlink{ref-latimerSurvivalAnalysisEconomic2013}{14}}
Censoring is common when collecting time-to-event data, since the follow-up time may not be long enough to observe the endpoint of interest for all individuals in the trial, and some individuals may be lost to follow-up.\textsuperscript{\protect\hyperlink{ref-collettModellingSurvivalData2015}{15}}
The likelihood function for survival data, \(\mathbf{x}\), obtained up until \(t_1\) for a model with hazard function \(h(\cdot)\) and survivor function \(S(\cdot)\) is
\begin{align}
\text{Likelihood } p(\mathbf{x}|\bm\theta) =\prod_{i=1}^{n_1}h(x_i, \bm\theta)^{\delta_i} S(x_i, \bm\theta),
\label{eq:likhood}
\end{align}

where \(i\) indexes the \(n_1=N\) study participants at risk at time zero, where the censoring indicator \(\delta_i=1\) when \(x_i\) is an observed event, \(\delta_i=0\) when \(x_i\) is a censored observation, and where \(\bm\theta\) are the parameters of the survival distribution.
The observed dataset at time point \(t_1\) consists of the \(n_1\) survival times and censoring indicators, \(\mathbf{x} = \{x_1, \ldots, x_{n_1}, \delta_1, \ldots, \delta_{n_1}\}\).

The data collected between time points \(t_1\) and \(t_2\) is denoted \(\mathbf{\tilde x} = \{\tilde x_{1}, \ldots, \tilde x_{n_2}, \tilde\delta_{1}, \ldots, \tilde\delta_{n_2}\}\), where \(n_2\) is the number of study participants at risk at \(t_1\).
The likelihood function for \(\mathbf{\tilde x}\) is left-truncated at \(t_1\) to reflect that events beyond \(t_1\) are conditional on not having occurred prior to \(t_1\).\textsuperscript{\protect\hyperlink{ref-kleinSurvivalAnalysisTechniques2013}{16}}
Unlike censoring, which contributes to the likelihood by plugging in a survival factor for censored observations as well as observed survival times, truncation does not add any data points to the likelihood.
This distinction is important, since we want to avoid double counting the observed data \(\mathbf{x}\) when we compute the likelihood for the ongoing study data \(\mathbf{\tilde{x}}\).
The left-truncated likelihood has an additional term in the denominator that re-normalises the truncated distribution so that it integrates to 1, i.e.

\begin{equation}
\text{Left-truncated likelihood } p_{LT}(\mathbf{\tilde{x}}|\bm\theta)=
\prod_{i=1}^{n_2}\frac{ h(\tilde x_i,\bm\theta)^{\tilde\delta_i}S(\tilde x_i,\bm\theta)}{S(t_1,\bm\theta)}.
\label{eq:likhood-trunc}    
\end{equation}

Once we have derived the posterior distribution for the model parameters given data at \(t_1\), \(p(\bm\theta|\mathbf{x})\), and the likelihood for the ongoing follow-up data, \(p_{LT}(\mathbf{\tilde{x}}|\bm\theta)\), we require a method for actually computing expression (\ref{eq:evsi_ongoing}).
In almost all realistic applications this will require numerical methods.
Nested Monte Carlo can be used, but this is computationally expensive.
A regression-based approach is much quicker,\textsuperscript{\protect\hyperlink{ref-strongEstimatingExpectedValue2015}{10}} and this is described along with the Monte Carlo approach in Appendix B.

We are now in a position to describe methods for computing EVSI that account for uncertainty about the choice of survival model.

\hypertarget{model-averaged-evsi-for-an-ongoing-study-accounting-for-survival-model-uncertainty}{%
\subsection{Model-averaged EVSI for an ongoing study accounting for survival model uncertainty}\label{model-averaged-evsi-for-an-ongoing-study-accounting-for-survival-model-uncertainty}}

\hypertarget{survival-model-uncertainty-and-model-averaging}{%
\subsubsection{Survival model uncertainty and model averaging}\label{survival-model-uncertainty-and-model-averaging}}

In this section, ``model'' refers to the survival model for the time-to-event data \(p(\mathbf{x}|\bm\theta)\), not the cost-effectiveness model, \(\NB(d, \bm\theta)\).
In many real applications we will be uncertain about which survival model is most appropriate and should be used to extrapolate the data beyond the observed follow-up period \(t_1\),
though we may be comfortable with proposing a candidate \textit{set} of models, \(\mathcal{M}=M_r,\; r=1, \dots,R\), that covers plausible approximations of the data generating process, i.e.~the set is \(\mathcal{M}-open\) in the terminology used by Bernardo and Smith (1994).\textsuperscript{\protect\hyperlink{ref-bernardoBayesianTheory1994}{17}}
In these circumstances, we may account for model uncertainty using \emph{predictive model averaging}, and average over model predictions using model weights based on each model's predictive ability.\textsuperscript{\protect\hyperlink{ref-jacksonAccountingUncertaintyHealth2009}{18},\protect\hyperlink{ref-jacksonStructuralParameterUncertainty2010}{19}}
After observing data \(\mathbf{x}\) at time \(t_1\), we place probability weight \(P(M_r|\mathbf{x})\) on the \(r^{th}\) model producing the best predictions, with \(\sum_{r=1}^R P(M_r|\mathbf{x}) = 1\).

The net benefit function for decision option \(d\) given model \(M_r\) and parameters \(\bm\theta_r\) is denoted \(\mathrm{NB}(d,\bm\theta_r, M_r)\).
Taking the expectation over both parameters and models after observing data \(\mathbf{x}\) up to time point \(t_1\) gives us
\begin{align}\nonumber
\text{Model-averaged NB}_d|\mathbf{x}&= \sum_{r=1}^{R}\left\{ \E_{\bm\theta_r|\mathbf{x},M_r}\mathrm{NB}(d,\bm\theta_r, M_r)P(M_r|\mathbf{x})\right\}\\\nonumber
& = \E_{\mathcal{M}|\mathbf{x}}[\E_{\theta_r|\mathbf{x},M_r}\{\mathrm{NB}(d,\bm\theta_r, M_r)\}]\\
& = \E_{\bm\theta_r,\mathcal{M}|\mathbf{x}}\{\mathrm{NB}(d,\bm\theta_r, M_r)\},
\label{eq:nb1-ma-1}
\end{align}
and the optimal choice at time point \(t_1\) is the decision \(d\) that maximises this expectation.

\hypertarget{evsi-for-an-ongoing-study-accounting-for-model-uncertainty}{%
\subsubsection{EVSI for an ongoing study accounting for model uncertainty}\label{evsi-for-an-ongoing-study-accounting-for-model-uncertainty}}

Additional follow-up data \(\mathbf{\tilde{x}}\) will not only update our judgements about parameters, \(p{(\bm\theta_r|\mathbf{x},\mathbf{\tilde{x}},M_r)}\), but will also update our judgements about the relative plausibility of each model, \(P{(M_r|\mathbf{x},\mathbf{\tilde{x}})}\), for each model \(r=1,\ldots,R\).

The EVSI for an ongoing study, where we average over models, is given by
\begin{align}
\text{Model-averaged } \EVSI
&=\E_{\mathbf{\tilde{x}}|\mathbf{x}} \Big[ \max_d \E_{\bm\theta_r, \mathcal{M}|\mathbf{x},\mathbf{\tilde{x}}} \{ \NB(d, \bm\theta_r, M_r) \} \Big] - \max_d \E_{\bm\theta_r, \mathcal{M}|\mathbf{x}} \{ \NB(d, \bm\theta_r, M_r) \},
\label{eq:evsi-ma-1}
\end{align}
which is identical to (\ref{eq:evsi_ongoing}), except that expectations are now taken over \textit{models} as well as parameters (see Appendix C for a derivation).

To compute (\ref{eq:evsi-ma-1}) we will need a method for generating plausible datasets \(\mathbf{\tilde{x}}\) from \(p(\mathbf{\tilde{x}}|\mathbf{x})\), the distribution of the follow-up data given the observed data, which takes account of the fact that we now consider plausible a number of different data generating models.
We will also need to define model probabilities given observed data, \(P(M_r|\mathbf{x})\), and then find a method for computing posterior model probabilities \(P(M_r|\mathbf{x},\mathbf{\tilde{x}})\), given each sampled future plausible dataset \(\mathbf{\tilde{x}}\).
We address the issue of defining model probabilities given observed data first.

\hypertarget{deriving-model-probabilities-given-observed-data-up-until-t_1}{%
\subsubsection{\texorpdfstring{Deriving model probabilities given observed data up until \(t_1\)}{Deriving model probabilities given observed data up until t\_1}}\label{deriving-model-probabilities-given-observed-data-up-until-t_1}}

We assume that before we see the observed data \(\mathbf{x}\), that we are indifferent about the `correct' model, so \(P(M_r) = \nicefrac{1}{R}\) for all \(r\).
After we observe data \(\mathbf{x}\), we use the Akaike's Information Criterion (AIC)\textsuperscript{\protect\hyperlink{ref-akaikeInformationTheoryExtension1973}{20}} to derive posterior model probabilities giving greater weight to models with better predictive ability (according to Kullback-Leibler divergence), as described by Jackson, Thompson, and Sharples (2009)\textsuperscript{\protect\hyperlink{ref-jacksonAccountingUncertaintyHealth2009}{18}}.
We set
\begin{align}
P(M_r|\mathbf{x}) &= \frac{\exp\{-0.5 \; \mathrm{AIC}_r(\mathbf{x})\}}{\sum_{r=1}^{R} \exp\{-0.5 \; \mathrm{AIC}_r(\mathbf{x})\}}, 
\label{eq:post-model}
\end{align}
where
\begin{align}
\mathrm{AIC}_r(\mathbf{x}) &= -2 \log\{p(\mathbf{x}| \hat{\bm\theta}_r)\} + 2u_r. \nonumber
\end{align}
The term \(\hat{\bm\theta}_r\) is the maximum likelihood estimate for the parameters of model \(M_r\), and \(u_r\) is the number of parameters in model \(M_r\).

\hypertarget{generating-plausible-ongoing-follow-up-datasets-mathbftildex-that-we-may-observe-between-t_1-and-t_2}{%
\subsubsection{\texorpdfstring{Generating plausible ongoing follow-up datasets, \(\mathbf{\tilde{x}}\), that we may observe between \(t_1\) and \(t_2\)}{Generating plausible ongoing follow-up datasets, \textbackslash mathbf\{\textbackslash tilde\{x\}\}, that we may observe between t\_1 and t\_2}}\label{generating-plausible-ongoing-follow-up-datasets-mathbftildex-that-we-may-observe-between-t_1-and-t_2}}

Plausible datasets from the distribution \(p(\mathbf{\tilde x}|\mathbf{x})\) are generated as follows.
Firstly, we sample a model \(M_r^{(k)}\) with probability \(P(M_r|\mathbf{x})\) given by Equation (\ref{eq:post-model}).
Next, we draw a sample \(\bm\theta_r^{(k)}\) from the distribution of the parameters of our chosen model \(p(\bm\theta_r|\mathbf{x}, M_r^{(k)})\).
Finally, we generate a dataset \(\mathbf{\tilde{x}}^{(k)}\) from the distribution of the data \(p(\mathbf{\tilde x}|\bm\theta_r^{(k)},M_r^{(k)})\) given the sampled parameter values \(\bm\theta_r^{(k)}\) and model \(M_r^{(k)}\). We can repeat this process \(k=1,\ldots,K\) times to generate an arbitrary number of datasets.

\hypertarget{updating-model-probabilities-given-ongoing-follow-up-data-from-t_1-to-t_2}{%
\subsubsection{\texorpdfstring{Updating model probabilities given ongoing follow-up data from \(t_1\) to \(t_2\)}{Updating model probabilities given ongoing follow-up data from t\_1 to t\_2}}\label{updating-model-probabilities-given-ongoing-follow-up-data-from-t_1-to-t_2}}

We can derive our posterior model probabilities at time point \(t_2\), for dataset \(\mathbf{\tilde{x}}^{(k)}\), via Bayes theorem:
\begin{equation}
P(M_r|\mathbf{x},\mathbf{\tilde{x}}^{(k)}) = \frac{p(\mathbf{\tilde{x}}^{(k)}|M_r, \mathbf{x})P(M_r|\mathbf{x})}{\sum_{r=1}^{R} p(\mathbf{\tilde{x}}^{(k)}|M_r,\mathbf{x})P(M_r|\mathbf{x})},
\label{eq:bayes}
\end{equation}
where \(p(\mathbf{\tilde{x}}^{(k)}|M_r,\mathbf{x})\) is the marginal likelihood (`marginal' because we have integrated out the model parameters):
\begin{equation}\nonumber
p(\mathbf{\tilde{x}}^{(k)}|M_r,\mathbf{x}) = \int_{\Theta} p(\mathbf{\tilde{x}}^{(k)}|M_r, \bm\theta_r) p(\bm\theta_r|M_r,\mathbf{x})\mathrm{d}\bm\theta_r.
\end{equation}

We use \textit{bridge sampling} to approximate the marginal likelihood, which is a form of importance sampling that has been shown to give good approximations in a wide range of settings.\textsuperscript{\protect\hyperlink{ref-mengSimulatingRatiosNormalizing1996}{21}--\protect\hyperlink{ref-wongPropertiesBridgeSampler2020}{24}}
The key notion behind bridge sampling is that the marginal likelihood can be written as the ratio of two expectations, each of which can be estimated via importance sampling.
The name `bridge' reflects the incorporation in the estimator of a density function that `bridges' (i.e.~has good overlap with) the two densities from which samples are drawn.
A detailed tutorial on the bridge sampling method is given in the article by Gronau et al.~(2017)\textsuperscript{\protect\hyperlink{ref-gronauTutorialBridgeSampling2017}{23}}, and the method is straightforward to implement in the R package \texttt{bridgesampling}\textsuperscript{\protect\hyperlink{ref-gronauBridgesamplingPackageEstimating2020}{25}}.
Given the bridge sampling estimates of \(p(\mathbf{\tilde{x}}^{(k)}|M_r,\mathbf{x})\) for each model, posterior model probabilities are trivial to compute via expression (\ref{eq:bayes}).

As with single-model EVSI, computing model-averaged EVSI (expression \ref{eq:evsi-ma-1}) will require numerical methods.
Nested Monte Carlo and a regression-based approach are described in Appendix D.
In the next section, we will apply these methods in a synthetic case study.

\hypertarget{synthetic-case-study}{%
\section{Synthetic case study}\label{synthetic-case-study}}

We will model survival with and without accounting for survival model uncertainty.

\hypertarget{decision-problem-and-model-definition-1}{%
\subsection{Decision problem and model definition}\label{decision-problem-and-model-definition-1}}

Our decision problem is to determine which of two treatment options has the longest mean survival; a new treatment \((d=1)\), or standard care \((d=2)\).

In the single-model case, survival is assumed to follow a Weibull distribution, and the net benefit of each treatment option is assumed to equal the restricted mean survival time, given an overall time horizon of \(t_h = 240\) months (i.e.~the area under the survival curve from 0 to 240 months).
So the net benefit function is:
\begin{equation}
\mathrm{NB}(d,\bm\theta_d)= \int_{0}^{t_h} \text{exp}\left\{-\biggl(\frac{t}{e^{\theta_{\lambda d}}}\biggr)^{e^{\theta_{kd}}}\right\} \mathrm{d}t,
\label{eq:nb1-weib}
\end{equation}
where the model parameters are the log-transformed Weibull shape and scale parameters, \(\bm\theta_d = (\theta_{kd}, \theta_{\lambda d})\).
Computing restricted mean survival for distributions other than the exponential requires numerical integration, but easy-to-use functions are available in the R package \texttt{flexsurv}.\textsuperscript{\protect\hyperlink{ref-jacksonFlexsurvPlatformParametric2016}{26}}

In the model-averaged case, the decision problem is as above, but we assume we are uncertain about the choice of survival model, \(M_r\), to extrapolate the observed data beyond the current follow-up period \(t_1\).
We assume that our set of plausible models \(\mathcal{M}\) contains the following four parametric distributions: Weibull \((r = 1)\), Gamma \((r = 2)\), Lognormal \((r = 3)\), and Log-logistic \((r = 4)\).

\hypertarget{generating-synthetic-case-study-datasets-mathbfx-collected-up-to-t_112-months}{%
\subsection{\texorpdfstring{Generating synthetic case study datasets, \(\mathbf{x}\), collected up to \(t_1=12\) months}{Generating synthetic case study datasets, \textbackslash mathbf\{x\}, collected up to t\_1=12 months}}\label{generating-synthetic-case-study-datasets-mathbfx-collected-up-to-t_112-months}}

We generated two synthetic case study datasets: one in which the hazard of death is monotonically increasing, and the other in which it is monotonically decreasing.
For each case study we generated a dataset with 200 participants per trial arm with a maximum follow-up of \(t_1=12\) months.
We denote the datasets \(\mathbf{x}_1\) for new treatment and \(\mathbf{x}_2\) for standard care.

To explore the performance of the method when the survival model was mis-specified we generated survival times evenly spaced from either a Weibull or a Gamma distribution, using the \(0.005^{th}, 0.015^{th}, \ldots, 0.985^{th}, 0.995^{th}\) quantiles from each distribution (i.e.~100 evenly spaced quantiles that avoid 0 and 1).
We could have randomly generated survival times, but this would have just added additional Monte Carlo error when assessing the methods for computing EVSI.
The parameters of the Weibull and Gamma distributions that we used to generate the synthetic case study datasets are shown in Table \ref{tab:hyperparams}.

\begin{table}

\begin{threeparttable}
\caption{\label{tab:hyperparams}Weibull and Gamma distribution parameters for the synthetic case study datasets}
\centering
\fontsize{9}{11}\selectfont
\begin{tabular}[t]{lcccc}
\toprule
\multicolumn{1}{c}{ } & \multicolumn{2}{c}{Increasing hazard case study} & \multicolumn{2}{c}{Decreasing hazard case study} \\
\cmidrule(l{3pt}r{3pt}){2-3} \cmidrule(l{3pt}r{3pt}){4-5}
  & New treatment & Standard care & New treatment & Standard care\\
\midrule
Weibull shape, $k$ & 1.10 & 1.10 & 0.60 & 0.60\\
Weibull scale, $\lambda$ & 70.00 & 50.00 & 80.00 & 57.00\\
Gamma shape, $\alpha$ & 1.80 & 1.80 & 0.80 & 0.80\\
Gamma rate, $\beta$ & 0.04 & 0.04 & 0.01 & 0.01\\
\bottomrule
\end{tabular}
\begin{tablenotes}
\small
\item []  
\end{tablenotes}
\end{threeparttable}
\end{table}

\begin{figure}

{\centering \includegraphics{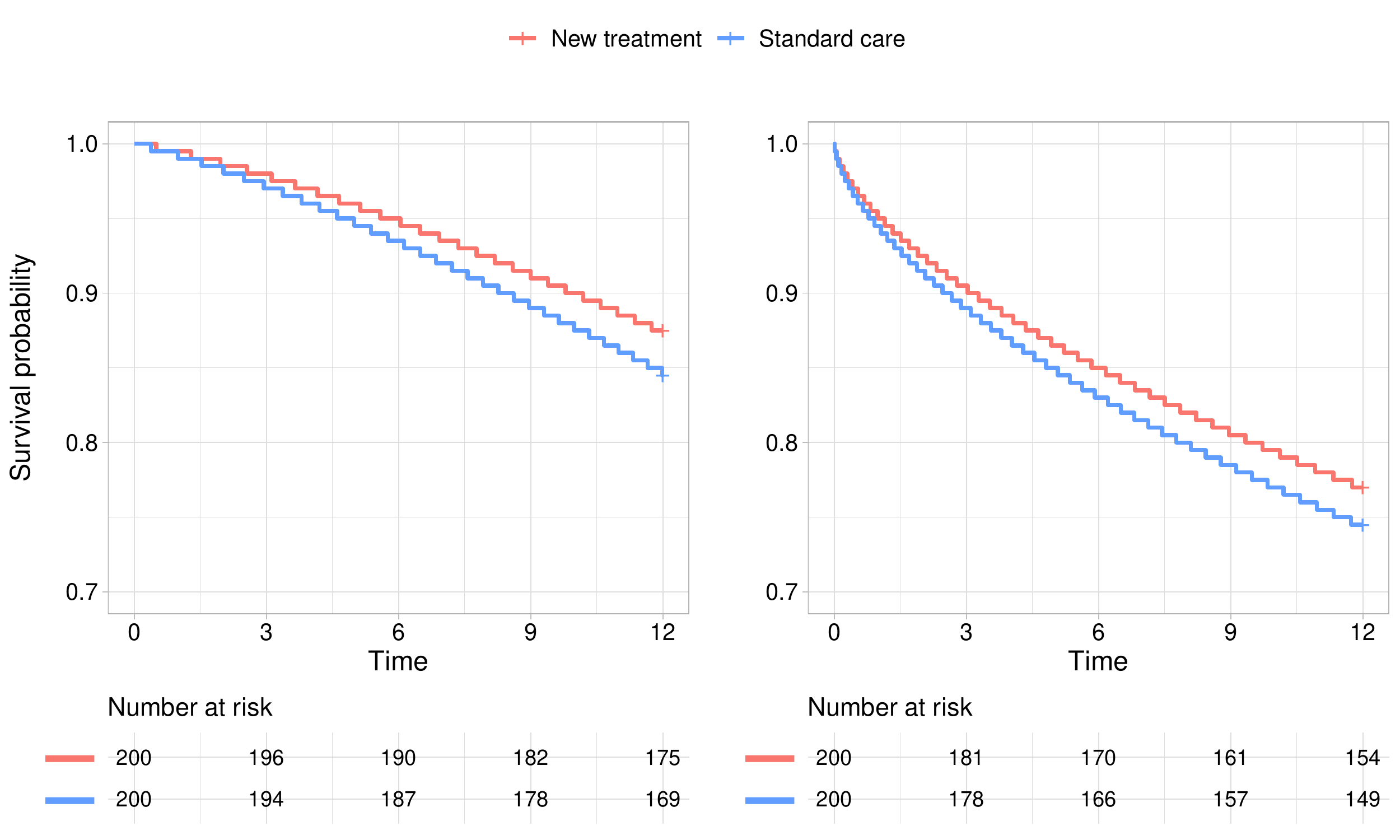} 

}

\caption{Kaplan-Meier plots for the increasing hazard dataset (left) and decreasing hazard dataset (right)}\label{fig:ipd-fig1}
\end{figure}

We enrolled all patients in the trial at \(t_0 = 0\), and right-censored the datasets at \(t_1 =\) 12 months.
We assumed no loss to follow-up and did not apply any other censoring.
Figure \ref{fig:ipd-fig1} shows Kaplan-Meier plots for the two synthetic case study datasets.

\hypertarget{initial-trial-analysis-at-t_112-months}{%
\subsection{\texorpdfstring{Initial trial analysis at \(t_1=12\) months}{Initial trial analysis at t\_1=12 months}}\label{initial-trial-analysis-at-t_112-months}}

For each synthetic case study, we analysed the two trial arms separately.
We fitted all four models to the data from each arm and estimated the model parameters using maximum likelihood (as implemented in the \texttt{flexsurvreg} function).\textsuperscript{\protect\hyperlink{ref-jacksonFlexsurvPlatformParametric2016}{26}}
We assumed that our judgements about the log-transformed parameters for each survival model conditional on the observed data up to \(t_1\), \(p(\bm\theta_r|\mathbf{x})\), are represented by a bivariate Normal distribution with the mean vector and covariance matrix derived from the maximum likelihood estimation.
We computed the AIC for each model fit and derived model probability weights via Equation (\ref{eq:post-model}).

Net benefits, AICs and model probabilities are shown in Table \ref{tab:ma-weights}, and means and covariances for each model are reported in Appendix G.

\begin{table}

\begin{threeparttable}
\caption{\label{tab:ma-weights}Mean survival, Akaike's Information Criterion and prior model probabilities \(P(M_r|\mathbf{x})\) for the two hypothetical datasets}
\centering
\fontsize{9}{11}\selectfont
\begin{tabular}[t]{lcccccc}
\toprule
\multicolumn{1}{c}{} & \multicolumn{3}{c}{Increasing hazard dataset} & \multicolumn{3}{c}{Decreasing hazard dataset} \\
\cmidrule(l{3pt}r{3pt}){2-4} \cmidrule(l{3pt}r{3pt}){5-7}
\multicolumn{1}{l}{ } & \multicolumn{1}{l}{\makecell[l]{Net benefit\\ (mean survival)}} & \multicolumn{1}{l}{$\text{AIC}(\mathbf{x})$} & \multicolumn{1}{l}{\(P(M_r|\mathbf{x})\)} & \multicolumn{1}{l}{\makecell[l]{Net benefit\\ (mean survival)}} & \multicolumn{1}{l}{$\text{AIC}(\mathbf{x})$} & \multicolumn{1}{l}{\(P(M_r|\mathbf{x})\)}\\
\midrule
\addlinespace[0.3em]
\multicolumn{7}{l}{\textit{New treatment}}\\
\hspace{1em}Weibull & 50.96 & 277.58 & 0.26 & 84.81 & 437.06 & 0.29\\
\hspace{1em}Gamma & 57.71 & 277.57 & 0.26 & 74.41 & 437.08 & 0.29\\
\hspace{1em}Lognormal & 110.43 & 277.97 & 0.22 & 123.49 & 438.45 & 0.14\\
\hspace{1em}Log-logistic & 79.28 & 277.58 & 0.26 & 105.98 & 437.14 & 0.28\\
\hspace{1em}\textbf{Weighted average} & \textbf{72.93} &  &  & \textbf{93.31} &  & \\
\addlinespace[0.3em]
\multicolumn{7}{l}{\textit{Standard care}}\\
\hspace{1em}Weibull & 44.01 & 329.26 & 0.28 & 77.85 & 470.37 & 0.30\\
\hspace{1em}Gamma & 49.42 & 329.29 & 0.28 & 66.99 & 470.40 & 0.29\\
\hspace{1em}Lognormal & 98.43 & 330.18 & 0.18 & 116.25 & 472.01 & 0.13\\
\hspace{1em}Log-logistic & 71.00 & 329.33 & 0.27 & 99.70 & 470.47 & 0.28\\
\hspace{1em}\textbf{Weighted average} & \textbf{62.36} &  &  & \textbf{85.85} &  & \\
\addlinespace[0.3em]
\multicolumn{7}{l}{\textit{Incremental values}}\\
\hspace{1em}\textbf{Weighted average} & \textbf{10.57} &  &  & \textbf{7.46} &  & \\
\bottomrule
\end{tabular}
\begin{tablenotes}
\small
\item [] AIC, Akaike’s Information Criterion.
\end{tablenotes}
\end{threeparttable}
\end{table}

The expected net benefits (mean survival times) assuming a single Weibull model computed via Equation (\ref{eq:nb1-weib}) are 50.96 versus 44.01 months (incremental = 6.95 months) for the increasing hazard dataset, and 84.81 versus 77.85 months (incremental = 6.97 months) for the decreasing hazard dataset.
The Expected Value of Perfect Information (EVPI) values, computed via Monte Carlo simulation with a sample size of \(10^5\), are 4.93 and 6.33 months for the increasing and decreasing hazard dataset, respectively.

The model-averaged net benefits, weighted by model probabilities, are 72.93 versus 62.36 months (incremental = 10.57 months) for the increasing hazard dataset, and 93.31 versus 85.85 months (incremental = 7.46 months) for the decreasing hazard dataset.
The model-averaged EVPI values are 10.32 and 9.97 months for the respective datasets.

\hypertarget{generating-plausible-ongoing-follow-up-datasets-mathbftildex-for-the-evsi-computation}{%
\subsection{\texorpdfstring{Generating plausible ongoing follow-up datasets, \(\mathbf{\tilde{x}}\), for the EVSI computation}{Generating plausible ongoing follow-up datasets, \textbackslash mathbf\{\textbackslash tilde\{x\}\}, for the EVSI computation}}\label{generating-plausible-ongoing-follow-up-datasets-mathbftildex-for-the-evsi-computation}}

Both the nested Monte Carlo and regression-based EVSI methods require a set of sampled ongoing follow-up datasets for each trial arm, denoted \(\mathbf{\tilde{x}}_1\) and \(\mathbf{\tilde{x}}_2\).
We generated \(k=1,\ldots,K\) datasets with \(K=\) 6,000 for each trial arm, where the \(k^{th}\) dataset was generated as follows.

In the single-model case, we first sampled log-shape and log-scale values, (\(\bm\theta_1^{(k)}\) for new treatment and \(\bm\theta_2^{(k)}\) for standard care),
from the bivariate Normal distributions in Appendix G.
We computed the net benefit for each decision option, given the sampled parameters, \(\NB(d, \bm\theta_d^{(k)})\) and stored this (these values are required for the regression-based approximation).
For each arm, we then sampled \(n\) survival times from a \textit{truncated} Weibull distribution (see Appendix E) with the sampled shape and scale values where \(n\) was the number of patients who were still alive in the trial arm at \(t_1=12\) months.
Finally, survival times were censored at the proposed endpoint for the ongoing data collection, \(t_2\).

In the model-averaged case, we first chose a model \(M_r^{(k)}\) with probability \(P(M_r|\mathbf{x})\), before sampling \(\bm\theta_r^{(k)}\) from the bivariate Normal distribution \(p(\bm\theta_r|\mathbf{x})\) for the chosen model \(M_r^{(k)}\) and generating the \(n\) survival times for each arm.
The remainder of the data generation step is as above.

\hypertarget{computing-evsi-for-ongoing-follow-up-via-nested-monte-carlo}{%
\subsection{Computing EVSI for ongoing follow-up via nested Monte Carlo}\label{computing-evsi-for-ongoing-follow-up-via-nested-monte-carlo}}

To sample from the posterior distributions, \(p(\bm\theta_d|\mathbf{x}_d,\mathbf{\tilde x}_d^{(k)})\), we used Hamiltonian Monte Carlo (HMC) as implemented in the package \texttt{rstan}\textsuperscript{\protect\hyperlink{ref-R-rstan}{27}}.
HMC is a Metropolis-Hastings MCMC algorithm with a particularly efficient sampling scheme that reduces Monte Carlo sampling error, therefore requiring fewer posterior samples for any inference.
The package \texttt{rstan} is an R interface to the Stan language.\textsuperscript{\protect\hyperlink{ref-gelmanStanProbabilisticProgramming2015}{28}}
An alternative option would have been to use OpenBUGS.\textsuperscript{\protect\hyperlink{ref-lunnBUGSProjectEvolution2009}{29}}

In the single-model case, for each outer loop sampled dataset, \(k = 1,\ldots,\) 6,000.00, we averaged the net benefit functions over \(J =\) 2,000.00 inner loop posterior samples of the model parameters, and stored the maximum net benefit of the two treatment options.
We then averaged these maximised net benefits and subtracted the expected value of a decision based on current information to obtain the EVSI following expression (\ref{eq:evsi-MC}) in Appendix B.

In the model-averaged case, for each outer loop dataset, we generated the \(J\) posterior samples of the model parameters for each of the \(r = 1,\ldots,4\) models (we needed to identify the truncated likelihood function for each model as we did for the Weibull example above, but this is straightforward. See Appendix E).
We weighted the parameter averaged net benefits \({\NB}_r^k(d)\) by the posterior model probabilities \(P(M_r | \mathbf{\tilde{x}}^{(k)})\) to give the posterior model-averaged expected net benefit, and identified the treatment \(d\) that maximized this for iteration \(k=1,\dots,\) 6,000.00.
We then subtracted the expected value of a decision based on current information to obtain the EVSI following expression (\ref{eq:evsi-ave-MC}) in Appendix D.

\hypertarget{computing-evsi-for-ongoing-follow-up-via-regression}{%
\subsection{Computing EVSI for ongoing follow-up via regression}\label{computing-evsi-for-ongoing-follow-up-via-regression}}

The GAM approach to computing EVSI for extending the follow-up until time \(t_2\) for the hypothetical example is as follows.

For each trial arm, we computed a low dimensional summary statistic for each dataset.
A convenient choice here is the number of observed events \(e_d^{(k)}\) and the total time at risk \(y_d^{(k)}\) for each dataset \(\mathbf{\tilde{x}}_d^{(k)}\), i.e.~\(T(\mathbf{\tilde{x}}_d^{(k)}) = \{e_d^{(k)},y_d^{(k)}\}\) for \(d=1,2\).

Then, for each of the two decision options, we fitted a GAM regression model with the stored net benefits \(\NB(d, \bm\theta_d^{(k)})\) as the dependent variable, and the two summary statistics, \(e_d^{(k)}\) and \(y_d^{(k)}\) as independent variables.
We allowed a smooth, arbitrary, non-linear relationship between the independent and dependent variables, plus arbitrary interaction between the independent variables, by specifying a `tensor product' cubic regression spline basis for the independent variables. This has the simple syntax \verb|gam(nb_d ~ te(e_d, y_d))| in the \texttt{mgcv}\textsuperscript{\protect\hyperlink{ref-woodMgcvMixedGAM2020}{30}} package in R.
We extracted the GAM model fitted values \(\hat{g}_{d}^{(k)}\) from each regression model fit, and estimated the EVSI using Equation \eqref{eq:gam-4} in Appendix B.

The GAM-based approximation method for model-averaged EVSI is identical to that used in the single-model case.

\hypertarget{results}{%
\section{Results}\label{results}}

\hypertarget{evsi-values-for-the-weibull-ongoing-data}{%
\subsection{EVSI values for the Weibull ongoing data}\label{evsi-values-for-the-weibull-ongoing-data}}

The nested Monte Carlo- and GAM-based EVSI estimates for additional follow-up times of 12, 24, 36, and 48 months (i.e.~\(t_2 = 24, 36, 48, 60\) months) are shown in Table \ref{tab:evsi-weib}.
The methods used to estimate the standard errors of the nested Monte Carlo and GAM estimators are described in an appendix of the article by Strong, Oakley, and Brennan (2014).\textsuperscript{\protect\hyperlink{ref-strongEstimatingMultiparameterPartial2014}{31}}

\begin{table}

\begin{threeparttable}
\caption{\label{tab:evsi-weib}EVSI (SE) values for additional follow-up time for the two hypothetical datasets given a Weibull distribution for the survival times}
\centering
\fontsize{9}{11}\selectfont
\begin{tabular}[t]{>{\raggedright\arraybackslash}p{3.4cm}>{\raggedright\arraybackslash}p{2.2cm}>{\raggedright\arraybackslash}p{2.2cm}>{\raggedright\arraybackslash}p{2.2cm}>{\raggedright\arraybackslash}p{2.2cm}}
\toprule
\multicolumn{1}{c}{ } & \multicolumn{2}{c}{Increasing hazard dataset} & \multicolumn{2}{c}{Decreasing hazard dataset} \\
\cmidrule(l{3pt}r{3pt}){2-3} \cmidrule(l{3pt}r{3pt}){4-5}
Additional follow-up (months) & Nested Monte Carlo & GAM & Nested Monte Carlo & GAM\\
\midrule
12 & 4.25 (0.09) & 4.28 (0.08) & 4.41 (0.10) & 4.46 (0.10)\\
24 & 4.58 (0.09) & 4.62 (0.06) & 5.20 (0.11) & 5.27 (0.09)\\
36 & 4.68 (0.09) & 4.71 (0.05) & 5.45 (0.11) & 5.54 (0.08)\\
48 & 4.74 (0.09) & 4.77 (0.04) & 5.55 (0.11) & 5.65 (0.07)\\
\bottomrule
\end{tabular}
\begin{tablenotes}
\small
\item [] GAM, Generalized Additive Model. EVPI values are 4.93 and 6.33, respectively. Total computation times for the analyses in the table are 24,808 seconds (Nested Monte Carlo) and 36 seconds (GAM).
\end{tablenotes}
\end{threeparttable}
\end{table}

As expected, the EVSI reflects diminishing marginal returns for increasing follow-up duration and converges towards the EVPI.
The EVSI varies depending on the underlying hazard pattern, even when point estimates of mean incremental survival benefit are similar (6.95 months for the increasing hazard dataset and 6.97 months for the decreasing hazard dataset).
The increasing hazard dataset has lower numbers of prior observed events and higher expected numbers of future events for the additional follow-up time than the decreasing hazard dataset, which - all else equal - is expected to result in greater EVSI values.
This upwards effect on EVSI is however canceled out by the downwards effect of lower estimates of mean survival, resulting in greater EVSI values for the decreasing hazard dataset than for the increasing hazard dataset.

The GAM method agrees well with the MCMC method, with the benefit of a greatly reduced computational cost.
The MCMC inner loop for the Monte Carlo method used parallel processing, but even with this additional efficiency, the regression method was approximately 700 times faster than the nested Monte Carlo method.
We used a machine running Windows 10 with an Intel\(\textsuperscript{\textregistered}\) Core\(\textsuperscript{TM}\) i9 CPU with 15 threads running on 8 cores at 2.40GHz, and with 32 GB RAM.

Of note is that the standard errors for the nested Monte Carlo estimator slightly increase with increasing follow-up duration, while the opposite is true for the GAM estimator.
This is due to different mechanisms through which the effective sample size of the generated data \(\mathbf{\tilde{x}}\) affects the standard errors of the nested Monte Carlo and GAM estimators, which is further explained in Appendix F.

\hypertarget{model-averaged-evsi-values}{%
\subsection{Model-averaged EVSI values}\label{model-averaged-evsi-values}}

The nested Monte Carlo- and GAM-based model-averaged EVSI estimates for additional follow-up times of 12, 24, 36, and 48 months (i.e.~\(t_2 = 24, 36, 48, 60\) months) are shown in Table \ref{tab:evsi-ma}.

\begin{table}

\begin{threeparttable}
\caption{\label{tab:evsi-ma}EVSI (SE) values for additional follow-up time for the two hypothetical datasets given a Weibull distribution for the survival times}
\centering
\fontsize{9}{11}\selectfont
\begin{tabular}[t]{>{\raggedright\arraybackslash}p{3.4cm}>{\raggedright\arraybackslash}p{2.2cm}>{\raggedright\arraybackslash}p{2.2cm}>{\raggedright\arraybackslash}p{2.2cm}>{\raggedright\arraybackslash}p{2.2cm}}
\toprule
\multicolumn{1}{c}{ } & \multicolumn{2}{c}{Increasing hazard dataset} & \multicolumn{2}{c}{Decreasing hazard dataset} \\
\cmidrule(l{3pt}r{3pt}){2-3} \cmidrule(l{3pt}r{3pt}){4-5}
\multicolumn{1}{>{\raggedright\arraybackslash}p{3.4cm}}{Additional follow-up (months)} & \multicolumn{1}{>{\raggedright\arraybackslash}p{2.2cm}}{Nested Monte Carlo} & \multicolumn{1}{>{\raggedright\arraybackslash}p{2.2cm}}{GAM} & \multicolumn{1}{>{\raggedright\arraybackslash}p{2.2cm}}{Nested Monte Carlo} & \multicolumn{1}{>{\raggedright\arraybackslash}p{2.2cm}}{GAM}\\
\midrule
12 & 7.50 (0.18) & 7.52 (0.14) & 6.69 (0.15) & 6.70 (0.13)\\
24 & 8.75 (0.20) & 8.82 (0.10) & 8.09 (0.18) & 8.16 (0.11)\\
36 & 9.43 (0.21) & 9.44 (0.08) & 8.71 (0.19) & 8.76 (0.09)\\
48 & 9.77 (0.22) & 9.74 (0.07) & 8.96 (0.19) & 9.01 (0.08)\\
\bottomrule
\end{tabular}
\begin{tablenotes}
\small
\item [] GAM, Generalized Additive Model. EVPI values are 10.32 and 9.97, respectively. Total computation times for the analyses in the table are 289,211 seconds (Nested Monte Carlo) and 37 seconds (GAM).
\end{tablenotes}
\end{threeparttable}
\end{table}

As expected, the EVSI converges towards the EVPI as follow-up time increases, and there is good agreement between the two methods.
The model-averaged EVSI values for additional follow-up are greater than the Weibull model EVSI (Table \ref{tab:evsi-weib}), which reflects the additional value in reducing model as well as parameter uncertainty.
The GAM method is approximately 8,000 times faster than the nested Monte Carlo method.

\hypertarget{expected-net-benefit-of-sampling}{%
\subsection{Expected Net Benefit of Sampling}\label{expected-net-benefit-of-sampling}}

The net value of additional data collection can be quantified by computing the Expected Net Benefit of Sampling (ENBS).\textsuperscript{\protect\hyperlink{ref-claxtonEconomicApproachClinical1996}{32}}
In the context of an ongoing study, the ENBS is the difference between the EVSI for collecting additional data between \(t_1\) and \(t_2\) and the expected cost of continuing the study and potential health benefits foregone if approval is withheld.
When the ENBS is positive, it is worthwhile to continue the study and collect more data before making an adoption decision.
If the adoption decision is reversible, approval can be granted while additional data is being collected.
This is referred to as ``approval with research'' (AWR).\textsuperscript{\protect\hyperlink{ref-claxtonComprehensiveAlgorithmApproval2016}{6}}
If the adoption decision is irreversible, approval should be withheld until the additional data has been collected, which is referred to as ``only in research'' (OIR).

\begin{figure}

{\centering \includegraphics{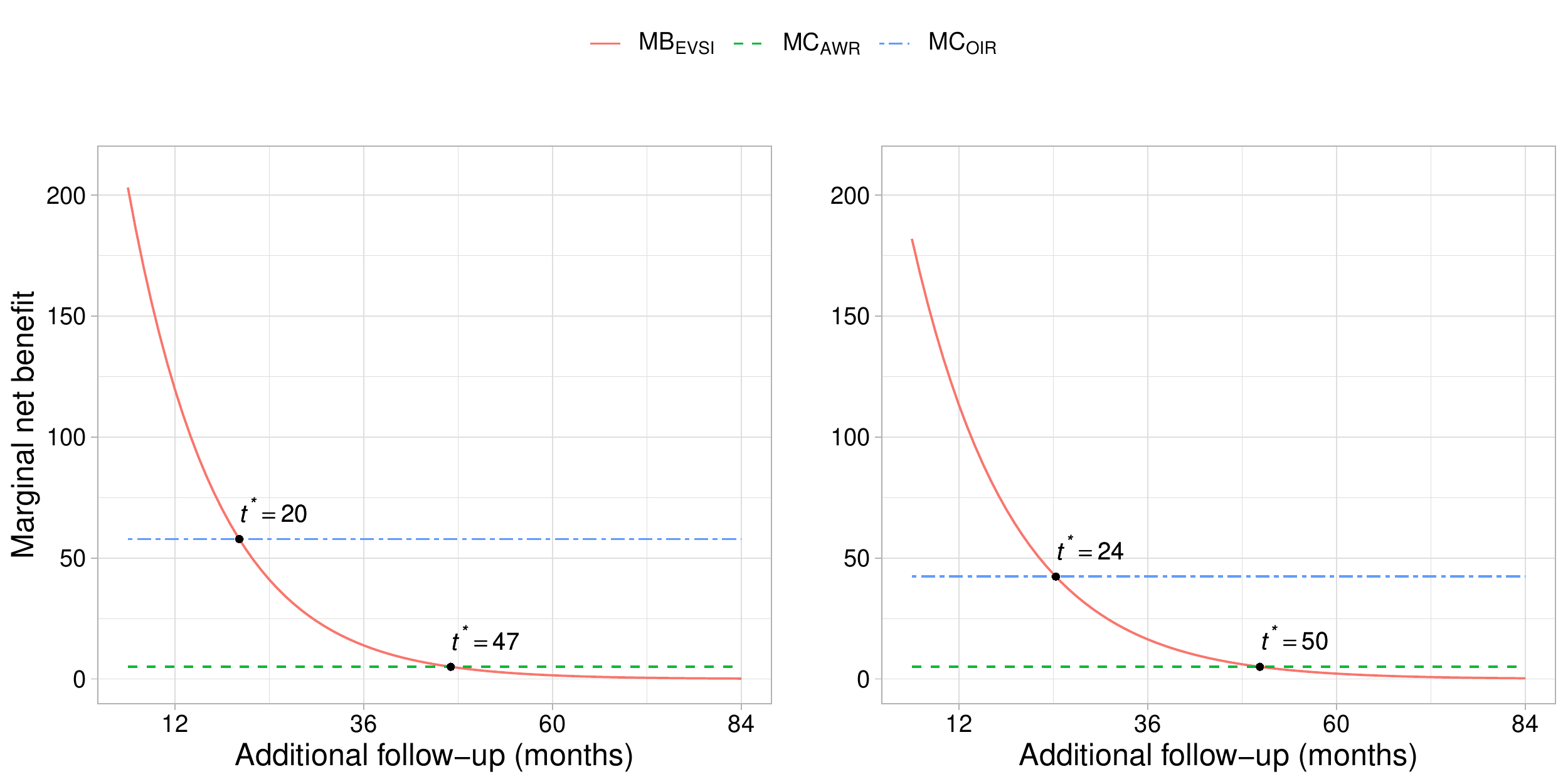} 

}

\caption{Marginal benefit (\(\text{MB}_{\text{EVSI}}\)), marginal cost of `approval with research' (\(\text{MC}_{\text{AWR}}\)) and marginal cost of `only in research' (\(\text{MC}_{\text{OIR}}\)) given different durations of additional follow-up. Estimates are based on the model-averaged EVSI analyses for the increasing hazard dataset (left) and decreasing hazard dataset (right), trial costs of 5 life months per month, 5 new patients receiving treatment each month and a decision time horizon of 10 years.}\label{fig:marg-evsi-fig}
\end{figure}

Figure \ref{fig:marg-evsi-fig} illustrates that when approval is reversible and AWR can be recommended, the marginal benefit in terms of model-averaged EVSI equals the marginal cost of continuing the trial at 47 and 50 months of additional follow-up for the increasing and decreasing hazard datasets, respectively.
These are the time points at which the ENBS is at a maximum.
When approval is irreversible and OIR is recommended, the ENBS is at a maximum when the marginal benefit of delaying the decision until more data has been collected equals the marginal cost of continuing the trial and withholding approval, which is at 20 and 24 months of additional follow-up for the increasing and decreasing hazard datasets, respectively.

\hypertarget{discussion}{%
\section{Discussion}\label{discussion}}

EVSI is useful not only for informing the design of a future trial, but also for deciding whether an ongoing study should continue in order to collect additional data before making an adoption decision.
This article is the first to set out generic EVSI algorithms for survival data from an ongoing trial with or without accounting for survival model uncertainty.
The EVSI algorithms generalise to any decision context in which structural uncertainty is present, provided that the analyst is able to derive probability weights for the competing scenarios.

\hypertarget{strengths-and-limitations}{%
\subsubsection{Strengths and limitations}\label{strengths-and-limitations}}

The nonparametric regression-based method is fast and straightforward to implement, even when we include consideration of model uncertainty.
In fact, extending the method to include model uncertainty does not increase the complexity or computation time.
The nested Monte Carlo procedure, on the other hand, is extremely computationally demanding when we include model uncertainty.

When a large part of the relevant time horizon is unobserved, the clinical plausibility of the survival extrapolations is often of greater importance than the mathematical fit to the observed data.\textsuperscript{\protect\hyperlink{ref-latimerSurvivalAnalysisEconomic2013}{14}}
Deriving prior model probabilities from purely statistical measures such as AIC may therefore not always be appropriate when data are immature, since these measures do not reflect the plausibility of the extrapolations.\textsuperscript{\protect\hyperlink{ref-gallacherExtrapolatingParametricSurvival2020}{8}}
This became evident in the hypothetical case studies, as the AIC-based prior model probabilities of the lognormal and log-logistic models were similar to those of the Weibull and Gamma models for the increasing hazard dataset, despite the fact that the former two models do not allow for monotonically increasing hazards and therefore cannot capture the true underlying hazard pattern.

An alternative approach to dealing with model uncertainty could be to consider a single very flexible model that includes all the models the analyst believes plausible.
For example, the Generalized F distribution includes most commonly used parametric survival distributions as special cases.\textsuperscript{\protect\hyperlink{ref-coxGeneralizedDistributionUmbrella2008}{33}}
It is however more common to view model uncertainty as structural uncertainty in choosing between competing survival models.\textsuperscript{\protect\hyperlink{ref-gallacherExtrapolatingParametricSurvival2021}{9},\protect\hyperlink{ref-jacksonFrameworkAddressingStructural2011}{34}}
Furthermore, the use of a very flexible model requires the specification of a prior that appropriately reflects uncertainty in choosing between alternative functional forms within the flexible model, which may be not be straightforward.
Flexible models such as the Generalized F distribution are also prone to overfitting and may not always provide reliable predictions of mean survival, particularly when data is immature.\textsuperscript{\protect\hyperlink{ref-gallacherExtrapolatingParametricSurvival2021}{9}}

Although we did not consider flexible parametric models such as Royston-Parmar spline-based models\textsuperscript{\protect\hyperlink{ref-roystonFlexibleParametricProportionalhazards2002}{35}} or mixture cure models\textsuperscript{\protect\hyperlink{ref-angelisMixtureModelsCancer1999}{36}} in our case studies, the principles outlined in this article apply to any parametric survival model.

In the synthetic case studies, we assumed all patients had the same follow-up at \(t_1\).
In clinical trials, patients are usually recruited over a period of time, which means the individual follow-up times will vary at \(t_1\).
In these circumstances, additional follow-up will not only provide more information about the tail of the survival curve (from patients that were enrolled early), but also about the central part (from patients that were enrolled later).

We did not consider sequential trial designs\textsuperscript{\protect\hyperlink{ref-armitageSearchOptimalityClinical1985}{37}}, which require EVSI to be recalculated after each observation and to account for all the possible ways in which future patients may be allocated to the trial arms or when to stop the trial.\textsuperscript{\protect\hyperlink{ref-briggsDecisionModellingHealth2006}{38}}
This can give rise to a large number of subproblems that may have to be solved using dynamic programming methods, which can be computationally very demanding.

\hypertarget{policy-implications}{%
\subsubsection{Policy implications}\label{policy-implications}}

Immature evidence leads to a high level of decision uncertainty, which may result in the uptake of technologies that reduce net health benefit.
The decision making context in which trials are ongoing and evidence is immature is particularly pronounced for new oncology drugs.
The purpose of the Cancer Drug Fund (CDF) in the UK, for example, is to enable early patient access to promising new cancer drugs while allowing evidential uncertainty to be reduced through ongoing data collection.
In the period between 2017 and July 2018, the National Institute for Health and Care Excellence (NICE) recommended over half of the appraised cancer drugs through the CDF, typically due to concerns about immature survival data.\textsuperscript{\protect\hyperlink{ref-waltonReviewIssuesAffecting2019a}{39}}

The EVSI algorithms in this article can help decision makers determine whether early patient access to a new technology can be justified on the basis of the current evidence or whether more mature evidence is needed.
The option to enroll more patients into an ongoing trial should also be considered if it has a positive net value.
Unlike most of the existing work on EVSI that primarily targets commissioners and funders of research, EVSI for ongoing trials also addresses the policy context of decision makers who do not have the remit to commission additional research.

\hypertarget{data-availability}{%
\subsubsection{Data Availability}\label{data-availability}}

The analysis code used in this study is available from Github at
\href{https://github.com/matverv/evsi-survival-ongoing-trial}{https://github.com/matverv/evsi-survival-ongoing-trial}.

\hypertarget{acknowledgements}{%
\subsubsection{Acknowledgements}\label{acknowledgements}}

We would like to thank researchers at the Centre for Health Economics in York for their comments on this work during a seminar in April, 2020.
Financial support for this study was provided entirely by a grant from the Norwegian Research Council through NordForsk (298854).
The funding agreement ensured the authors' independence in designing the study, interpreting the data, writing, and publishing the report.

\hypertarget{declaration-of-conflicting-interests}{%
\subsubsection{Declaration of Conflicting Interests}\label{declaration-of-conflicting-interests}}

The authors have no conflicts of interest that are directly relevant to the content of this article.

\newpage

\hypertarget{references}{%
\section*{References}\label{references}}
\addcontentsline{toc}{section}{References}

\hypertarget{refs}{}
\begin{CSLReferences}{0}{0}
\leavevmode\hypertarget{ref-schlaiferProbabilityStatisticsBusiness1959}{}%
\CSLLeftMargin{1. }
\CSLRightInline{Schlaifer R. Probability and {Statistics} for {Business Decisions}. First Edition edition. {McGraw-Hill}; 1959. }

\leavevmode\hypertarget{ref-raiffaAppliedStatisticalDecision1961a}{}%
\CSLLeftMargin{2. }
\CSLRightInline{Raiffa H, Schlaifer R. Applied statistical decision theory. {Boston}: {Division of Research, Graduate School of Business Adminitration, Harvard University}; 1961. }

\leavevmode\hypertarget{ref-flightReviewClinicalTrials2019}{}%
\CSLLeftMargin{3. }
\CSLRightInline{Flight L, Arshad F, Barnsley R, Patel K, Julious S, Brennan A, Todd S. A {Review} of {Clinical Trials With} an {Adaptive Design} and {Health Economic Analysis}. Value in Health. 2019 Apr;22(4):391--398. }

\leavevmode\hypertarget{ref-europeanmedicinesagencyAdaptivePathways2018}{}%
\CSLLeftMargin{4. }
\CSLRightInline{European Medicines Agency. Adaptive pathways {[}Internet{]}. 2018 {[}cited 2020 Oct 19{]}. Available from: \url{https://www.ema.europa.eu/en/human-regulatory/research-development/adaptive-pathways}}

\leavevmode\hypertarget{ref-europeanmedicinesagencyConditionalMarketingAuthorisation2018}{}%
\CSLLeftMargin{5. }
\CSLRightInline{European Medicines Agency. Conditional marketing authorisation {[}Internet{]}. 2018 {[}cited 2020 Oct 19{]}. Available from: \url{https://www.ema.europa.eu/en/human-regulatory/marketing-authorisation/conditional-marketing-authorisation}}

\leavevmode\hypertarget{ref-claxtonComprehensiveAlgorithmApproval2016}{}%
\CSLLeftMargin{6. }
\CSLRightInline{Claxton K, Palmer S, Longworth L, Bojke L, Griffin S, Soares M, Spackman E, Rothery C. A {Comprehensive Algorithm} for {Approval} of {Health Technologies With}, {Without}, or {Only} in {Research}: {The Key Principles} for {Informing Coverage Decisions}. Value in Health. 2016 Sep;19(6):885--891. }

\leavevmode\hypertarget{ref-eckermannOptionValueDelay2008}{}%
\CSLLeftMargin{7. }
\CSLRightInline{Eckermann S, Willan AR. The {Option Value} of {Delay} in {Health Technology Assessment}. Medical Decision Making. 2008 May;28(3):300--305. }

\leavevmode\hypertarget{ref-gallacherExtrapolatingParametricSurvival2020}{}%
\CSLLeftMargin{8. }
\CSLRightInline{Gallacher D, Kimani P, Stallard N. Extrapolating {Parametric Survival Models} in {Health Technology Assessment}: {A Simulation Study}. Medical Decision Making. {SAGE Publications Inc STM}; 2020 Dec;0272989X20973201. }

\leavevmode\hypertarget{ref-gallacherExtrapolatingParametricSurvival2021}{}%
\CSLLeftMargin{9. }
\CSLRightInline{Gallacher D, Kimani P, Stallard N. Extrapolating {Parametric Survival Models} in {Health Technology Assessment Using Model Averaging}: {A Simulation Study}. Medical Decision Making. {SAGE Publications Inc STM}; 2021 Feb;0272989X21992297. }

\leavevmode\hypertarget{ref-strongEstimatingExpectedValue2015}{}%
\CSLLeftMargin{10. }
\CSLRightInline{Strong M, Oakley JE, Brennan A, Breeze P. Estimating the {Expected Value} of {Sample Information Using} the {Probabilistic Sensitivity Analysis Sample}: {A Fast}, {Nonparametric Regression}-{Based Method}. Medical Decision Making. 2015 Jul;35(5):570--583. }

\leavevmode\hypertarget{ref-heathCalculatingExpectedValue2020}{}%
\CSLLeftMargin{11. }
\CSLRightInline{Heath A, Kunst N, Jackson C, Strong M, Alarid-Escudero F, Goldhaber-Fiebert JD, Baio G, Menzies NA, Jalal H. Calculating the {Expected Value} of {Sample Information} in {Practice}: {Considerations} from 3 {Case Studies}. Medical Decision Making. {SAGE Publications Inc STM}; 2020 Apr;40(3):314--326. }

\leavevmode\hypertarget{ref-adesExpectedValueSample2004}{}%
\CSLLeftMargin{12. }
\CSLRightInline{Ades AE, Lu G, Claxton K. Expected {Value} of {Sample Information Calculations} in {Medical Decision Modeling}. Medical Decision Making. 2004 Mar;24(2):207--227. }

\leavevmode\hypertarget{ref-albertBayesianComputation2009}{}%
\CSLLeftMargin{13. }
\CSLRightInline{Albert J. Bayesian computation with {R}. 2. Aufl. {Dordrecht}: {Springer}; 2009. }

\leavevmode\hypertarget{ref-latimerSurvivalAnalysisEconomic2013}{}%
\CSLLeftMargin{14. }
\CSLRightInline{Latimer NR. Survival {Analysis} for {Economic Evaluations Alongside Clinical Trials} - {Extrapolation} with {Patient}-{Level Data}: {Inconsistencies}, {Limitations}, and a {Practical Guide}. Medical Decision Making. 2013 Aug;33(6):743--754. }

\leavevmode\hypertarget{ref-collettModellingSurvivalData2015}{}%
\CSLLeftMargin{15. }
\CSLRightInline{Collett D. Modelling {Survival Data} in {Medical Research}. 3 edition. {Chapman and Hall/CRC}; 2015. }

\leavevmode\hypertarget{ref-kleinSurvivalAnalysisTechniques2013}{}%
\CSLLeftMargin{16. }
\CSLRightInline{Klein JP, Moeschberger ML. Survival {Analysis}: {Techniques} for {Censored} and {Truncated Data}. {Springer Science \& Business Media}; 2013. }

\leavevmode\hypertarget{ref-bernardoBayesianTheory1994}{}%
\CSLLeftMargin{17. }
\CSLRightInline{Bernardo JM, Smith AFM. Bayesian {Theory}. {Chichester}: {Wiley}; 1994. }

\leavevmode\hypertarget{ref-jacksonAccountingUncertaintyHealth2009}{}%
\CSLLeftMargin{18. }
\CSLRightInline{Jackson CH, Thompson SG, Sharples LD. Accounting for uncertainty in health economic decision models by using model averaging. Journal of the Royal Statistical Society: Series A (Statistics in Society). 2009 Apr;172(2):383--404. }

\leavevmode\hypertarget{ref-jacksonStructuralParameterUncertainty2010}{}%
\CSLLeftMargin{19. }
\CSLRightInline{Jackson CH, Sharples LD, Thompson SG. Structural and parameter uncertainty in {Bayesian} cost-effectiveness models. Journal of the Royal Statistical Society: Series C (Applied Statistics). 2010 Mar;59(2):233--253. }

\leavevmode\hypertarget{ref-akaikeInformationTheoryExtension1973}{}%
\CSLLeftMargin{20. }
\CSLRightInline{Akaike H. Information {Theory} and an {Extension} of the {Maximum Likelihood Principle}. {In B}. {N}. {Petrov}, \& {F}. {Csaki} ({Eds}.), {Proceedings} of the 2nd {International Symposium} on {Information Theory} (pp. 267-281). 1973; }

\leavevmode\hypertarget{ref-mengSimulatingRatiosNormalizing1996}{}%
\CSLLeftMargin{21. }
\CSLRightInline{Meng X-L, Wong WH. Simulating {Ratios} of {Normalizing Constants} via a {Simple Identity}: {A Theoretical Exploration}. Statistica Sinica. {Institute of Statistical Science, Academia Sinica}; 1996;6(4):831--860. }

\leavevmode\hypertarget{ref-fruhwirth-schnatterEstimatingMarginalLikelihoods2004}{}%
\CSLLeftMargin{22. }
\CSLRightInline{Frühwirth-Schnatter S. Estimating marginal likelihoods for mixture and {Markov} switching models using bridge sampling techniques. The Econometrics Journal. 2004 Jun;7(1):143--167. }

\leavevmode\hypertarget{ref-gronauTutorialBridgeSampling2017}{}%
\CSLLeftMargin{23. }
\CSLRightInline{Gronau QF, Sarafoglou A, Matzke D, Ly A, Boehm U, Marsman M, Leslie DS, Forster JJ, Wagenmakers E-J, Steingroever H. A tutorial on bridge sampling. Journal of Mathematical Psychology. 2017 Dec;81:80--97. }

\leavevmode\hypertarget{ref-wongPropertiesBridgeSampler2020}{}%
\CSLLeftMargin{24. }
\CSLRightInline{Wong JST, Forster JJ, Smith PWF. Properties of the bridge sampler with a focus on splitting the {MCMC} sample. Statistics and Computing. 2020 Jul;30(4):799--816. }

\leavevmode\hypertarget{ref-gronauBridgesamplingPackageEstimating2020}{}%
\CSLLeftMargin{25. }
\CSLRightInline{Gronau QF, Singmann H, Wagenmakers E-J. Bridgesampling: {An R Package} for {Estimating Normalizing Constants}. Journal of Statistical Software. 2020 Feb;92(1):1--29. }

\leavevmode\hypertarget{ref-jacksonFlexsurvPlatformParametric2016}{}%
\CSLLeftMargin{26. }
\CSLRightInline{Jackson C. Flexsurv: {A Platform} for {Parametric Survival Modeling} in {R}. Journal of Statistical Software. 2016 May;70(1):1--33. }

\leavevmode\hypertarget{ref-R-rstan}{}%
\CSLLeftMargin{27. }
\CSLRightInline{Guo J, Gabry J, Goodrich B, Weber S. \texttt{Rstan}: R interface to stan {[}Internet{]}. 2020. Available from: \url{https://CRAN.R-project.org/package=rstan}}

\leavevmode\hypertarget{ref-gelmanStanProbabilisticProgramming2015}{}%
\CSLLeftMargin{28. }
\CSLRightInline{Gelman A, Lee D, Guo J. Stan: {A Probabilistic Programming Language} for {Bayesian Inference} and {Optimization}. Journal of Educational and Behavioral Statistics. 2015 Oct;40(5):530--543. }

\leavevmode\hypertarget{ref-lunnBUGSProjectEvolution2009}{}%
\CSLLeftMargin{29. }
\CSLRightInline{Lunn D, Spiegelhalter D, Thomas A, Best N. The {BUGS} project: {Evolution}, critique and future directions. Statistics in Medicine. 2009 Nov;28(25):3049--3067. PMID: \href{https://www.ncbi.nlm.nih.gov/pubmed/19630097}{19630097}}

\leavevmode\hypertarget{ref-woodMgcvMixedGAM2020}{}%
\CSLLeftMargin{30. }
\CSLRightInline{Wood S. \texttt{Mgcv}: {Mixed GAM Computation Vehicle} with {Automatic Smoothness Estimation} {[}Internet{]}. 2020. Available from: \url{https://CRAN.R-project.org/package=mgcv}}

\leavevmode\hypertarget{ref-strongEstimatingMultiparameterPartial2014}{}%
\CSLLeftMargin{31. }
\CSLRightInline{Strong M, Oakley JE, Brennan A. Estimating {Multiparameter Partial Expected Value} of {Perfect Information} from a {Probabilistic Sensitivity Analysis Sample}: {A Nonparametric Regression Approach}. Medical Decision Making. 2014 Apr;34(3):311--326. }

\leavevmode\hypertarget{ref-claxtonEconomicApproachClinical1996}{}%
\CSLLeftMargin{32. }
\CSLRightInline{Claxton K, Posnett J. An economic approach to clinical trial design and research priority-setting. Health Economics. 1996 Nov;5(6):513--524. }

\leavevmode\hypertarget{ref-coxGeneralizedDistributionUmbrella2008}{}%
\CSLLeftMargin{33. }
\CSLRightInline{Cox C. The generalized {F} distribution: {An} umbrella for parametric survival analysis. Statistics in Medicine. 2008;27(21):4301--4312. }

\leavevmode\hypertarget{ref-jacksonFrameworkAddressingStructural2011}{}%
\CSLLeftMargin{34. }
\CSLRightInline{Jackson CH, Bojke L, Thompson SG, Claxton K, Sharples LD. A {Framework} for {Addressing Structural Uncertainty} in {Decision Models}. Russell LB, editor. Medical Decision Making. 2011 Jul;31(4):662--674. }

\leavevmode\hypertarget{ref-roystonFlexibleParametricProportionalhazards2002}{}%
\CSLLeftMargin{35. }
\CSLRightInline{Royston P, Parmar MKB. Flexible parametric proportional-hazards and proportional-odds models for censored survival data, with application to prognostic modelling and estimation of treatment effects. Statistics in Medicine. 2002 Aug;21(15):2175--2197. PMID: \href{https://www.ncbi.nlm.nih.gov/pubmed/12210632}{12210632}}

\leavevmode\hypertarget{ref-angelisMixtureModelsCancer1999}{}%
\CSLLeftMargin{36. }
\CSLRightInline{Angelis RD, Capocaccia R, Hakulinen T, Soderman B, Verdecchia A. Mixture models for cancer survival analysis: Application to population-based data with covariates. Statistics in Medicine. 1999;18(4):441--454. }

\leavevmode\hypertarget{ref-armitageSearchOptimalityClinical1985}{}%
\CSLLeftMargin{37. }
\CSLRightInline{Armitage P. The {Search} for {Optimality} in {Clinical Trials}. International Statistical Review / Revue Internationale de Statistique. {{[}Wiley, International Statistical Institute (ISI){]}}; 1985;53(1):15--24. }

\leavevmode\hypertarget{ref-briggsDecisionModellingHealth2006}{}%
\CSLLeftMargin{38. }
\CSLRightInline{Briggs A, Claxton K, Sculpher M. Decision {Modelling} for {Health Economic Evaluation}. 1 edition. {Oxford}: {Oxford University Press}; 2006. }

\leavevmode\hypertarget{ref-waltonReviewIssuesAffecting2019a}{}%
\CSLLeftMargin{39. }
\CSLRightInline{Walton MJ, O'Connor J, Carroll C, Claxton L, Hodgson R. A {Review} of {Issues Affecting} the {Efficiency} of {Decision Making} in the {NICE Single Technology Appraisal Process}. PharmacoEconomics - Open. 2019 Sep;3(3):403--410. }

\leavevmode\hypertarget{ref-jalalGaussianApproximationApproach2018}{}%
\CSLLeftMargin{40. }
\CSLRightInline{Jalal H, Alarid-Escudero F. A {Gaussian Approximation Approach} for {Value} of {Information Analysis}. Medical Decision Making. 2018 Feb;38(2):174--188. }

\end{CSLReferences}

\newpage

\hypertarget{appendices}{%
\section*{Appendices}\label{appendices}}
\addcontentsline{toc}{section}{Appendices}

\beginappendixA

\hypertarget{appendix-a---deriving-an-expression-for-the-evsi-for-an-ongoing-study-assuming-no-model-uncertainty}{%
\subsection*{Appendix A - Deriving an expression for the EVSI for an ongoing study assuming no model uncertainty}\label{appendix-a---deriving-an-expression-for-the-evsi-for-an-ongoing-study-assuming-no-model-uncertainty}}
\addcontentsline{toc}{subsection}{Appendix A - Deriving an expression for the EVSI for an ongoing study assuming no model uncertainty}

Before the study starts, we have only prior knowledge about model parameters, which we represent via the distribution \(p(\bm\theta)\).
In many cases we will not have strong prior information, and \(p(\bm\theta)\) will therefore be minimally informative (typically flat on some scale).

We collect data \(\mathbf{x}\) during an initial period of follow-up that extends up until time \(t_1\).
At \(t_1\) we update our judgements about \(\bm\theta\), conditional on \(\mathbf{x}\) to give the posterior distribution \(p(\bm\theta|\mathbf{x})\).
The question is, should we continue the study to collect more data before making an adoption decision?

The optimal decision option given observed data up to time \(t_1\) has expected value
\begin{equation}\label{eq:max_pre_ongoing}
\max_d\E_{\bm\theta|\mathbf{x}} \{ \NB(d, \bm\theta) \}.
\end{equation}
Further data collection up until time point \(t_2\) will give us additional data \(\mathbf{\tilde{x}}\), which we can use to update judgements about \(\theta\) to give \(p(\bm\theta|\mathbf{x},\mathbf{\tilde{x}})\).
The optimum decision option will have expected value,
\begin{equation}\label{eq:max_post_ongoing}
\max_d\E_{\bm\theta|\mathbf{x},\mathbf{\tilde{x}}} \{ \NB(d, \bm\theta) \}.
\end{equation}
At time point \(t_1\) data \(\mathbf{\tilde{x}}\) are as yet uncollected, however we can take the expectation of expression
(\ref{eq:max_post_ongoing}) with respect to the distribution of the ongoing follow-up data \(\mathbf{\tilde{x}}\) conditional on the observed follow-up data \(\mathbf{x}\), \(p(\mathbf{\tilde{x}}|\mathbf{x})\), giving
\begin{equation}\label{eq:ex_max_post_ongoing}
\E_{\mathbf{\tilde{x}}|\mathbf{x}}[\max_d\E_{\bm\theta|\mathbf{x},\mathbf{\tilde{x}}} \{ \NB(d, \bm\theta) \}].
\end{equation}
The EVSI for continuing the study from \(t_1\) to \(t_2\) is the difference between the expected value of a decision made after collecting data up to \(t_2\), expression (\ref{eq:ex_max_post_ongoing}), and the expected value of a decision based on observed data collected up to \(t_1\), expression (\ref{eq:max_pre_ongoing}),
\begin{equation}
\EVSI\mathrm{(ongoing\;study)}=\E_{\mathbf{\tilde{x}}|\mathbf{x}}[\max_d\E_{\bm\theta|\mathbf{x},\mathbf{\tilde{x}}} \{ \NB(d, \bm\theta) \}] - \max_d\E_{\bm\theta|\mathbf{x}} \{ \NB(d, \bm\theta) \}.
\end{equation}

\newpage
\beginappendixB

\hypertarget{appendix-b---methods-for-computing-the-evsi-for-an-ongoing-study-assuming-no-model-uncertainty}{%
\subsection*{Appendix B - Methods for computing the EVSI for an ongoing study assuming no model uncertainty}\label{appendix-b---methods-for-computing-the-evsi-for-an-ongoing-study-assuming-no-model-uncertainty}}
\addcontentsline{toc}{subsection}{Appendix B - Methods for computing the EVSI for an ongoing study assuming no model uncertainty}

\hypertarget{nested-monte-carlo-method-for-computing-evsi-for-an-ongoing-study}{%
\subsubsection*{Nested Monte Carlo method for computing EVSI for an ongoing study}\label{nested-monte-carlo-method-for-computing-evsi-for-an-ongoing-study}}
\addcontentsline{toc}{subsubsection}{Nested Monte Carlo method for computing EVSI for an ongoing study}

Calculating EVSI for an ongoing study requires evaluation of the expectation of a maximised conditional expectation, \(\E_{\mathbf{\tilde{x}}|\mathbf{x}}[\max_d\E_{\theta|\mathbf{x},\mathbf{\tilde{x}}} \{ \NB(d, \bm\theta) \}]\).
This will rarely have an analytic solution. A nested expectation can be evaluated using a nested `double-loop' Monte Carlo scheme, which leads us to the following estimator for EVSI,
\begin{equation}
\EVSI\simeq \frac{1}{K} \sum_{k=1}^K \max_d \frac{1}{J} \sum_{j=1}^J \NB(d, \bm\theta^{(j,k)})  - \max_d \frac{1}{K} \sum_{k=1}^K  \frac{1}{J} \sum_{j=1}^J \NB(d, \bm\theta^{(j,k)}).
\label{eq:evsi-MC}
\end{equation}
In this scheme, we generate samples from \(p(\mathbf{\tilde x}|\mathbf{x})\) in the `outer loop.'
We do this by first sampling \(\bm\theta^{(k)},\, k=1,\ldots, K\) from \(p(\bm\theta|\mathbf{x})\), and then sampling \(\mathbf{\tilde x}^{(k)}\) from the truncated likelihood \(p_{LT}(\mathbf{\tilde x}|\bm\theta^{(k)})\).
For each sample \(\mathbf{\tilde x}^{(k)}\), we then sample values \(\bm\theta^{(j,k)},\, j=1,\ldots,J\) from the posterior distribution \(p(\bm\theta|\mathbf{x},\mathbf{\tilde x}^{(k)})\) in the `inner loop'.
Unless \(p(_{LT}\mathbf{\tilde x}|\bm\theta)\) and \(p(\bm\theta|\mathbf{x})\) are conjugate, which will be rare in practice, then sampling from \(p(\bm\theta|\mathbf{x},\mathbf{\tilde x}^{(k)})\) will require Markov Chain Monte Carlo (MCMC) or a similar scheme.
The total number of samples required for each \(d\) is \(J \times K\).

Note that the second term in expression (\ref{eq:evsi-MC}) has a nested double loop structure, even though the target estimand is the single maximised expectation \(\max_d\E_{\bm\theta|\mathbf{x}} \{ \NB(d, \bm\theta) \}\).
We reuse the same samples for both terms in the EVSI expression in order to reduce Monte Carlo error, noting that \(\max_d\E_{\mathbf{\tilde{x}}|\mathbf{x}}[\E_{\bm\theta|\mathbf{x},\mathbf{\tilde{x}}} \{ \NB(d, \bm\theta) \}] =\max_d\E_{\bm\theta|\mathbf{x}} \{ \NB(d, \bm\theta) \}\) by the law of total expectation.\textsuperscript{\protect\hyperlink{ref-strongEstimatingExpectedValue2015}{10}}

\hypertarget{regression-based-method-for-computing-evsi-for-an-ongoing-study}{%
\subsubsection*{Regression-based method for computing EVSI for an ongoing study}\label{regression-based-method-for-computing-evsi-for-an-ongoing-study}}
\addcontentsline{toc}{subsubsection}{Regression-based method for computing EVSI for an ongoing study}

Strong and others (2015)\textsuperscript{\protect\hyperlink{ref-strongEstimatingExpectedValue2015}{10}} developed a fast, non-parametric regression-based method that greatly reduces the computational burden of the nested Monte Carlo procedure to EVSI.
Their approach relies on estimating the functional relationship between the posterior expected net benefits and the generated datasets, thereby avoiding the inner loop and markedly increasing efficiency over the nested Monte Carlo method.

In the regression approach, we first generate a random parameter vector \(\bm\theta^{(k)}\) from the distribution of model parameters \(p(\bm\theta|\mathbf{x})\) at time point \(t_1\), and a random data sample \(\mathbf{\tilde{x}}^{(k)}\) from the truncated likelihood \(p_{LT}(\mathbf{\tilde{x}}|\bm\theta^{(k)})\), where \(k\) indicates the \(k_\text{th}\) sample.
The net benefit is evaluated at the same \(k^{th}\) sample of the model parameters, \(\NB(d,\bm\theta^{(k)})\).
We then express the observed net benefit \(\NB(d,\bm\theta^{(k)})\) as a sum of the conditional expectation of the net benefit given the data, \(\E_{\bm\theta|{\mathbf{x},\mathbf{\tilde{x}}^{(k)}}} \{ \NB(d,\bm\theta) \}\), which we wish to estimate to evaluate the EVSI (Equation (\ref{eq:evsi_ongoing})), and a mean-zero error term, \(\varepsilon^{(k)}\),
\begin{align}
\NB(d,\bm\theta^{(k)}) = \E_{\bm\theta|{\mathbf{x},\mathbf{\tilde{x}}^{(k)}}} \{ \NB(d,\bm\theta) \} + \varepsilon^{(k)}.         
\label{eq:gam-1}
\end{align}

As explained by Strong and others (2015)\textsuperscript{\protect\hyperlink{ref-strongEstimatingExpectedValue2015}{10}}, we can think of the conditional expectation \(\E_{\bm\theta|{\mathbf{x},\mathbf{\tilde{x}}^{(k)}}} \{ \NB(d,\bm\theta) \}\) as an unknown function of \(\mathbf{\tilde{x}}^{(k)}\).
We denote this function \(g(d, \mathbf{\tilde{x}}^{(k)})\) and substitute this into Equation \eqref{eq:gam-1}, giving
\begin{align}
\NB(d,\bm\theta^{(k)}) = g(d, \mathbf{\tilde{x}}^{(k)}) + \varepsilon^{(k)}.        
\label{eq:gam-2}
\end{align}
Since \(\mathbf{\tilde{x}}\) is a vector of (possibly censored) time-to-event data, and therefore high-dimensional, we write the the function \(g\) in terms of a low-dimensional summary statistic of the data \(T(\mathbf{\tilde{x}})\),
\begin{align}
\NB(d,\bm\theta^{(k)}) = g\{d, T(\mathbf{\tilde{x}}^{(k)}) \} + \varepsilon^{(k)}.    
\label{eq:gam-3}
\end{align}

We then use a generalized additive model (GAM), which is a flexible non-parametric regression model, to estimate the target function \(g\).
This means that we fit a GAM model to each decision option \(d\) and extract the regression model fitted values to estimate posterior net benefit.
We denote the GAM model fitted values as \(\hat{g}_{d}^{(k)}\).
The GAM-based EVSI estimate is given by
\begin{align}
\text{EVSI} \simeq \frac{1}{K} \sum_{k=1}^{K} \max_d \hat{g}_{d}^{(k)} - \max_d \frac{1}{K} \sum_{k=1}^{K} \hat{g}_{d}^{(k)}.     
\label{eq:gam-4}
\end{align}

\newpage
\beginappendixC

\hypertarget{appendix-c---deriving-an-expression-for-the-evsi-for-an-ongoing-study-accounting-for-model-uncertainty}{%
\subsection*{Appendix C - Deriving an expression for the EVSI for an ongoing study accounting for model uncertainty}\label{appendix-c---deriving-an-expression-for-the-evsi-for-an-ongoing-study-accounting-for-model-uncertainty}}
\addcontentsline{toc}{subsection}{Appendix C - Deriving an expression for the EVSI for an ongoing study accounting for model uncertainty}

In the model averaging setting, additional follow-up data \(\mathbf{\tilde{x}}\) will update our judgements about both parameters and the relative plausibility of each model.

The net benefit function for decision option \(d\) given model \(M_r\) and parameters \(\bm\theta_r\) is denoted \(\mathrm{NB}(d,\bm\theta_r, M_r)\).
At time point \(t_1\) after observing data \(\mathbf{x}\), the expected net benefit, averaging over both parameters and models is
\begin{align}\nonumber
\text{Model-averaged NB}_d|\mathbf{x}&= \sum_{r=1}^{R}\left\{ \E_{\bm\theta_r|\mathbf{x},M_r}\mathrm{NB}(d,\bm\theta_r, M_r)P(M_r|\mathbf{x})\right\}\\\nonumber
& = \E_{\mathcal{M}|\mathbf{x}}[\E_{\theta_r|\mathbf{x},M_r}\{\mathrm{NB}(d,\bm\theta_r, M_r)\}]\\
& = \E_{\bm\theta_r,\mathcal{M}|\mathbf{x}}\{\mathrm{NB}(d,\bm\theta_r, M_r)\},
\label{eq:nb1-ma-1a}
\end{align}
and the optimal choice at time point \(t_1\) is the decision \(d\) that maximises this expectation.

The net benefit after observing ongoing follow-up data \(\mathbf{\tilde x}\) between \(t_1\) and \(t_2\) is
\begin{align}\nonumber
\text{Model-averaged NB}_d|\mathbf{x},\mathbf{\tilde x}&= \sum_{r=1}^{m}\left\{ \E_{\bm\theta_r|\mathbf{x},\mathbf{\tilde x},M_r}\mathrm{NB}(d,\bm\theta_r, M_r) P(M_r|\mathbf{x},\mathbf{\tilde x})\right\}\\
& = \E_{\bm\theta_r,\mathcal{M}|\mathbf{x},\mathbf{\tilde x}}\{\mathrm{NB}(d,\bm\theta_r, M_r)\},
\label{eq:nb1-ma-2}
\end{align}
and the optimal choice at time point \(t_2\) is the decision \(d\) that maximises this expectation.
Follow-up data \(\mathbf{\tilde{x}}\) are not available at \(t_1\), but we can compute the \textit{expected value} of the maximised net benefit based on our beliefs from the data collected by \(t_1\),
\begin{equation}
\E_{\mathbf{\tilde{x}}|\mathbf{x}} \left[\max_d\E_{\bm\theta_r,\mathcal{M}|\mathbf{x},\mathbf{\tilde x}}\{\mathrm{NB}(d,\bm\theta_r, M_r)\}\right].
\label{eq:nb1-ma-3}
\end{equation}

The EVSI for an ongoing study, where we average over models, is then the difference between (\ref{eq:nb1-ma-3}) and the maximised value of (\ref{eq:nb1-ma-1a}),
\begin{align}
\text{Model-averaged } \EVSI
&=\E_{\mathbf{\tilde{x}}|\mathbf{x}} \Big[ \max_d \E_{\bm\theta_r, \mathcal{M}|\mathbf{x},\mathbf{\tilde{x}}} \{ \NB(d, \bm\theta_r, M_r) \} \Big] - \max_d \E_{\bm\theta_r, \mathcal{M}|\mathbf{x}} \{ \NB(d, \bm\theta_r, M_r) \}.
\end{align}

\newpage
\beginappendixD

\hypertarget{appendix-d---methods-for-computing-model-averaged-evsi-for-an-ongoing-study}{%
\subsection*{Appendix D - Methods for computing model-averaged EVSI for an ongoing study}\label{appendix-d---methods-for-computing-model-averaged-evsi-for-an-ongoing-study}}
\addcontentsline{toc}{subsection}{Appendix D - Methods for computing model-averaged EVSI for an ongoing study}

\hypertarget{nested-monte-carlo-method-for-computing-model-averaged-evsi}{%
\subsubsection*{Nested Monte Carlo method for computing model-averaged EVSI}\label{nested-monte-carlo-method-for-computing-model-averaged-evsi}}
\addcontentsline{toc}{subsubsection}{Nested Monte Carlo method for computing model-averaged EVSI}

The nested double-loop Monte Carlo scheme in expression (\ref{eq:evsi-MC}) naturally extends to the nested triple loop scheme when we average over models as well as over parameters and datasets,
\begin{align}\nonumber
\text{Model-averaged } \EVSI\simeq &\frac{1}{K} \sum_{k=1}^K \max_d  \sum_{r=1}^R  \frac{1}{J}\sum_{j=1}^J \NB(d, \bm\theta_r^{(j,k)}, M_r^{(k)})P(M_r^{(k)}|\mathbf{x},\mathbf{\tilde x}^{(k)})\\
& \; - \max_d \frac{1}{K} \sum_{k=1}^K \sum_{r=1}^R \frac{1}{J}\sum_{j=1}^J \NB(d, \bm\theta_r^{(j,k)}, M_r^{(k)})P(M_r^{(k)}|\mathbf{x},\mathbf{\tilde x}^{(k)}).
\label{eq:evsi-ave-MC}
\end{align}
In this (somewhat intimidating looking) scheme, we first generate \(k=1,\ldots,K\) samples \(\mathbf{\tilde x}^{(k)}\) from \(p(\mathbf{\tilde x}|\mathbf{x})\) in the `outer loop' (as described in the generating datasets section above).
Then, in the inner loop, we compute posterior expected net benefits by drawing \(j = 1,\dots,J\) samples \(\bm\theta_r^{(j,k)}\) from \(p{(\bm\theta_r|\mathbf{x},\mathbf{\tilde{x}}^{(k)},M_r^{(k)})}\) and take the average for each treatment \(d\).
This inner loop sampling from the posterior distribution of the parameters typically requires MCMC, unless the prior and truncated likelihood are conjugate.
Finally, for each \(k\), we compute the posterior model probability \(P(M_r^{(k)}|\mathbf{x},\mathbf{\tilde x}^{(k)})\) for each model \(r=1,\ldots,R\) (again, as described above).

As before, we reuse the same samples for both terms in the EVSI expression in order to reduce Monte Carlo error, noting that \(\max_d\E_{\mathbf{\tilde{x}}|\mathbf{x}}(\E_{\mathcal{M}|\mathbf{x},\mathbf{\tilde{x}}} [\E_{\bm\theta_r|\mathbf{x},\mathbf{\tilde{x}},M_r} \{ \NB(d, \bm\theta_r,M_r) \}] )=\max_d\E_{\bm\theta_r,\mathcal{M}|\mathbf{x}} \{ \NB(d, \bm\theta_r,M_r) \}\) by the law of total expectation.
The total number of samples required for each \(d\) is \(J \times R \times K\).

The nested triple-loop Monte Carlo scheme for computing model-averaged EVSI is given in Box \ref{alg:mcmc-evsi}.

\begin{algorithm}
\SetAlgoLined
\For{$k=1,\dots,K$ \textup{outer loops}}{
Sample a model \(M_r^{(k)}\) given current data $\mathbf x$ with probability \(P(M_r|\mathbf{x})\)\\
Sample \(\bm\theta_r^{(k)}\) from the distribution of the parameters of the sampled model, \(p(\bm\theta_r|\mathbf{x}, M_r^{(k)})\)\\
Generate a new data sample \(\mathbf{\tilde{x}}^{(k)}\) from the distribution of the data \(p(\mathbf{\tilde x}|\bm\theta_r^{(k)},M_r^{(k)})\)\\

  \For{$r=1,\dots,R$ \textup{models}}{
  Compute posterior expected net benefits by drawing \(j = 1,\dots,J\) inner loop samples \(\bm\theta_r^{(j,k)}\) from \(p{(\bm\theta_r|\mathbf{x},\mathbf{\tilde{x}}^{(k)},M_r^{(k)})}\) and take the average for each decision option \(d\)
  }
  Compute the posterior model probabilities \(P(M_1^{(k)},\dots, M_R^{(k)}|\mathbf{x},\mathbf{\tilde x}^{(k)})\)\\
  Find the decision option $d$ that maximises model-averaged posterior expected net benefit for iteration $k$\\
}
Compute the expected value of a decision based on new data $\mathbf{\tilde{x}}$ by taking the average of the maximum expected net benefits over the $K$ iterations\\
Compute the expected value of a decision based on current data $\mathbf{x}$ by finding the decision option $d$ that maximises the average of the expected net benefits over the $K$ iterations\\
Compute the EVSI by subtracting the expected value of a decision based on current data from the expected value of a decision based on new data
\caption{Nested Monte Carlo Scheme for Computing Model-Averaged EVSI}
\label{alg:mcmc-evsi}
\end{algorithm}

\hypertarget{regression-based-method-for-computing-model-averaged-evsi}{%
\subsubsection*{Regression-based method for computing model-averaged EVSI}\label{regression-based-method-for-computing-model-averaged-evsi}}
\addcontentsline{toc}{subsubsection}{Regression-based method for computing model-averaged EVSI}

The non-parametric regression-based method for computing model-averaged EVSI is a natural extension of the regression-based method for a single known model described above.
Firstly, we sample a model \(M_r^{(k)}\) with probability \(P(M_r|\mathbf{x})\) given by Equation (\ref{eq:post-model}).
Next, we draw a sample \(\bm\theta_r^{(k)}\) from the distribution of the parameters of our chosen model \(p(\bm\theta_r|\mathbf{x}, M_r^{(k)})\).

We then generate a dataset \(\mathbf{\tilde{x}}^{(k)}\) from the distribution of the data \(p(\mathbf{\tilde x}|\bm\theta_r^{(k)},M_r^{(k)})\) given the sampled parameter values \(\bm\theta_r^{(k)}\) and model \(M_r^{(k)}\).
Finally, we compute the net benefit, \(\NB(d,\bm\theta_r^{(k)},M_r^{(k)})\) for each \(d\). Repeating this \(k=1,\ldots,K\) times gives us, for each \(d\), a vector of \(K\) net benefits, and \(K\) corresponding datasets \(\mathbf{\tilde{x}}^{(1)},\ldots,\mathbf{\tilde{x}}^{(K)}\).

We express (for each \(d\)) the observed model-averaged net benefit \(\NB(d,\bm\theta_r^{(k)}, M_r^{(k)})\) as a sum of the posterior expectation of the net benefit given dataset \(\mathbf{\tilde{x}}^{(k)}\) and a mean-zero error term,
\begin{align}
\NB(d,\bm\theta_r^{(k)}, M_r^{(k)}) = \E_{\bm\theta_r, \mathcal{M}|\mathbf{x},\mathbf{\tilde{x}}^{(k)}} \{ \NB(d,\bm\theta_r,M_r) \} + \varepsilon^{(k)}.       
\label{eq:gam-ma-1}
\end{align}
We can think of the expectation \(\E_{\bm\theta_r, \mathcal{M}|\mathbf{x},\mathbf{\tilde{x}}^{(k)}} \{ \NB(d,\bm\theta_r, M_r) \}\) as an unknown function of \(\mathbf{\tilde{x}}^{(k)}\), which we denote \(g(d,\mathbf{\tilde{x}}^{(k)})\).
Substituting this into Equation (\ref{eq:gam-ma-1}) gives
\begin{align}
\NB(d,\bm\theta_r^{(k)}, M_r^{(k)}) = g(d, \mathbf{\tilde{x}}^{(k)}) + \varepsilon^{(k)}.        
\label{eq:gam-ma-2}
\end{align}
This means that the posterior model-averaged net benefit can be expressed in terms of a single function \(g\) and error term \(\varepsilon\), independent of the number of models \(m\) considered in the analysis.
We write the the function \(g\) in terms of a low-dimensional summary statistic of the survival data \(T(\mathbf{\tilde{x}}^{(k)})\),
\begin{align}
\NB(d,\bm\theta_r^{(k)}, M_r^{(k)}) = g\{d, T(\mathbf{\tilde{x}}^{(k)}) \} + \varepsilon^{(k)}.        
\label{eq:gam-ma-3}
\end{align}
We then estimate the posterior model-averaged net benefit as before, by fitting a GAM model to each decision option \(d\) and extracting the regression model fitted values \(\hat{g}_{d}^{(k)}\).
The model-averaged EVSI is then given by Equation (\ref{eq:gam-4}).

The GAM regression-based scheme for computing model-averaged EVSI is given in Box \ref{alg:gam-evsi}.

\begin{algorithm}
\SetAlgoLined
\For{$k=1,\dots,K$ \textup{outer loops}}{
Sample a model \(M_r^{(k)}\) given current data $\mathbf x$ with probability \(P(M_r|\mathbf{x})\)\\
Sample \(\bm\theta_r^{(k)}\) from the distribution of the parameters of the sampled model, \(p(\bm\theta_r|\mathbf{x}, M_r^{(k)})\)\\
Evaluate net benefit \(\NB(d,\bm\theta_r^{(k)},M_r^{(k)})\)\\
Generate a new data sample \(\mathbf{\tilde{x}}^{(k)}\) from the distribution of the data \(p(\mathbf{\tilde x}|\bm\theta_r^{(k)},M_r^{(k)})\)\\
Calculate a summary statistic \(T(\mathbf{\tilde{x}}^{(k)})\)
  }
Regress the net benefits \(\NB(d,\bm\theta_r^{(k)},M_r^{(k)})\) on \(T(\mathbf{\tilde{x}}^{(k)})\) for each decision option $d$ using GAM\\
Extract the GAM fitted values \(\hat{g}_{d}^{(k)}\) for each $d$\\
Compute EVSI via Equation \eqref{eq:gam-4}
\caption{Generalized Additive Model (GAM) Regression-Based Scheme for Computing Model-Averaged EVSI}
\label{alg:gam-evsi}
\end{algorithm}

\newpage
\beginappendixE

\hypertarget{appendix-e---truncated-likelihood-functions-for-the-weibull-gamma-lognormal-and-log-logistic-survival-models}{%
\subsection*{Appendix E - Truncated likelihood functions for the Weibull, Gamma, Lognormal and Log-logistic survival models}\label{appendix-e---truncated-likelihood-functions-for-the-weibull-gamma-lognormal-and-log-logistic-survival-models}}
\addcontentsline{toc}{subsection}{Appendix E - Truncated likelihood functions for the Weibull, Gamma, Lognormal and Log-logistic survival models}

Note that we use the same parameterisations of the Weibull, Gamma, Lognormal and Log-logistic distributions as in the R package \texttt{flexsurv}.\textsuperscript{\protect\hyperlink{ref-jacksonFlexsurvPlatformParametric2016}{26}}

\hypertarget{truncated-likelihood-function-for-the-weibull-distribution}{%
\subsubsection*{Truncated likelihood function for the Weibull distribution}\label{truncated-likelihood-function-for-the-weibull-distribution}}
\addcontentsline{toc}{subsubsection}{Truncated likelihood function for the Weibull distribution}

In order to compute EVSI via the nested Monte Carlo scheme described by Equation (\ref{eq:evsi-MC}) we need to define the Weibull truncated likelihood functions for the generated data: \(p(\mathbf{\tilde{x}}_1^{(k)}|\theta_{k1},\theta_{\lambda 1})\) for new treatment and \(p(\mathbf{\tilde{x}}_2^{(k)}|\theta_{k2},\theta_{\lambda 2})\) for standard care.

The Weibull hazard function for the new treatment arm given log-shape \(\theta_{k 1}\), log-scale \(\theta_{\lambda 1}\) and survival time \(x\) is
\begin{equation}
h(x, \theta) = \frac{e^{\theta_{k1}}}{e^{\theta_{\lambda 1}}}\left(\frac{x}{e^{\theta_{\lambda 1}}}\right)^{e^{\theta_{k1}}-1}.
\end{equation}
The survivor function is
\begin{equation}
S(x, \theta) = e^{-\left( {x}/{e^{\theta_{\lambda 1}}}\right)^{e^{\theta_{k1}}}},
\end{equation}
and the left-truncated likelihood function is therefore
\begin{align}
\text{Left-truncated likelihood } p(\mathbf{\tilde  x}|\theta_{k1}, \theta_{\lambda 1})
&=\prod_{i=1}^{n_2}\left[ \frac{\left\{\frac{e^{\theta_{k1}}}{e^{\theta_{\lambda 1}}}\left(\frac{\tilde x_i}{e^{\theta_{\lambda 1}}}\right)^{e^{\theta_{k1}}-1}\right\}^{\tilde \delta_i} e^{-\left({\tilde x_i}/{e^{\theta_{\lambda 1}}}\right)^{e^{\theta_{k1}}}}}
{e^{-\left({t_1}/{e^{\theta_{\lambda 1}}}\right)^{e^{\theta_{k1}}}}}\right].
\end{align}
where \(\tilde{x}_i\) and \(\tilde{\delta}_i\) are the survival time and censoring indicator for patient \(i\), where censoring is at the proposed new follow-up time of \(t_2\).
The expressions above are similarly defined for standard care \((d=2)\) with \(\theta_{k2},\theta_{\lambda 2}\) replacing \(\theta_{k1},\theta_{\lambda 1}\), and \(\mathbf{\tilde x}_2\) replacing \(\mathbf{\tilde x}_1\).

Let \(i\) index the \(n_1=N\) study participants at risk at time zero, where the censoring indicator \(\delta_i=1\) when \(x_i\) is an observed event, \(\delta_i=0\) when \(x_i\) is a censored observation, and where \(\bm\theta\) are the parameters of the survival distribution.
The observed dataset at time point \(t_1\) consists of the \(n_1\) survival times and censoring indicators, \(\mathbf{x} = \{x_1, \ldots, x_{n_1}, \delta_1, \ldots, \delta_{n_1}\}\).
Denote the data collected between time points \(t_1\) and \(t_2\) as \(\mathbf{\tilde x} = \{\tilde x_{1}, \ldots, \tilde x_{n_2}, \tilde\delta_{1}, \ldots, \tilde\delta_{n_2}\}\), where \(n_2\) is the number of study participants at risk at \(t_1\).
Events occurring between \(t_1\) and \(t_2\) are conditional on not having occurred before \(t_1\).

\hypertarget{truncated-likelihood-function-for-the-gamma-distribution}{%
\subsubsection*{Truncated likelihood function for the Gamma distribution}\label{truncated-likelihood-function-for-the-gamma-distribution}}
\addcontentsline{toc}{subsubsection}{Truncated likelihood function for the Gamma distribution}

The Gamma density function given log-shape \(\theta_{\alpha}\), log-rate \(\theta_{\beta}\) and survival time \(x\) is

\begin{equation*}
f(x, \theta) = \frac{(e^{\theta_{\beta}})^{e^{\theta_{\alpha}}}} {{\Gamma(e^{\theta_{\alpha}})}} x^{e^{\theta_{\alpha}}-1} e^{-x e^{\theta_{\beta}}}.
\end{equation*}

The survivor function is
\begin{equation*}
S(x, \theta) = 1-\gamma(e^{\theta_{\alpha}},x),
\end{equation*}
where \(\gamma(e^{\theta_{\alpha}},x)\) is the lower incomplete gamma function, given by
\begin{equation*}
\gamma(e^{\theta_{\alpha}},x) = \frac{1}{\Gamma(e^{\theta_{\alpha}})} \int_0^{x} u^{e^{\theta_{\alpha}}-1} e^{-u} \text{d}u.
\end{equation*}

We can define the left-truncated likelihood function for the Gamma distribution in terms of the density function and survivor function,
\begin{align*}
\text{Left-truncated likelihood } p(\mathbf{\tilde  x}|\theta_{\alpha}, \theta_{\beta}) &= \prod_{i=1}^{n_2}  \left[\frac{ f(\tilde x_i,\theta_{\alpha}, \theta_{\beta})^{\tilde\delta_i}S(\tilde x_i,\theta_{\alpha}, \theta_{\beta})^{1-\tilde\delta_i}}{S(t_1,\theta_{\alpha}, \theta_{\beta})}\right] \\
&= \prod_{i=1}^{n_2} \left[\frac{\left\{ \frac{(e^{\theta_{\beta}})^{e^{\theta_{\alpha}}}} {{\Gamma(e^{\theta_{\alpha}})}} \tilde x_i^{e^{\theta_{\alpha}}-1} e^{-\tilde x_i e^{\theta_{\beta}}} \right\}^{\tilde \delta_i} \left\{1-\gamma(e^{\theta_{\alpha}},\tilde x_i)\right\}^{1-\tilde \delta_i}}{1- \gamma(e^{\theta_{\alpha}},t_1)} \right].
\end{align*}

\hypertarget{truncated-likelihood-function-for-the-lognormal-distribution}{%
\subsubsection*{Truncated likelihood function for the Lognormal distribution}\label{truncated-likelihood-function-for-the-lognormal-distribution}}
\addcontentsline{toc}{subsubsection}{Truncated likelihood function for the Lognormal distribution}

The Lognormal density function given mean \(\theta_{\mu}\) and log-standard deviation \(\theta_{\sigma}\) of the logarithm, and survival time \(x\) is
\begin{equation*}
f(x, \theta) = \frac{1}{x e^{\theta_\sigma} \sqrt{2\pi}} e^{-\left(  \frac{(\log x - \theta_\mu)^2}{2 (e^{\theta_\sigma})^2}  \right)}.
\end{equation*}

The survivor function is
\begin{equation*}
S(x, \theta) = 1-\Phi \left( \frac{\log x - \theta_{\mu}}{e^{\theta_{\sigma}}}  \right),
\end{equation*}
where \(\Phi\) is the cumulative distribution function of the standard normal distribution \(\mathcal{N}(0,1)\).

The hazard function is given by
\begin{align*}
h(x, \theta) &= \frac{f(x, \theta)}{S(x, \theta)}\\
&=\frac{\frac{1}{x e^{\theta_\sigma} \sqrt{2\pi}} e^{-\left(  \frac{(\log x - \theta_\mu)^2}{2 (e^{\theta_\sigma})^2}  \right)}}{1-\Phi \left( \frac{\log x - \theta_{\mu}}{e^{\theta_{\sigma}}}  \right)}
\end{align*}

and the left-truncated likelihood function is

\begin{align*}
\text{Left-truncated likelihood } p(\mathbf{\tilde  x}|\theta_{\mu}, \theta_{\sigma})
&=\prod_{i=1}^{n_2} \left[ \frac{                     
\left\{ \frac{\frac{1}{\tilde x_i e^{\theta_\sigma} \sqrt{2\pi}} e^{-\left(  \frac{(\log \tilde x_i - \theta_\mu)^2}{2 (e^{\theta_\sigma})^2}  \right)}}{1-\Phi \left( \frac{\log \tilde x_i - \theta_{\mu}}{e^{\theta_{\sigma}}}\right)}\right\}^{\tilde\delta_i} \left\{ 1-\Phi \left( \frac{\log \tilde x_i - \theta_{\mu}}{e^{\theta_{\sigma}}} \right) \right\}
}{1-\Phi \left( \frac{\log t_1 - \theta_{\mu}}{e^{\theta_{\sigma}}}  \right)}     \right].
\label{eq:likhood-lnorm-trunc}
\end{align*}

\hypertarget{truncated-likelihood-function-for-the-log-logistic-distribution}{%
\subsubsection*{Truncated likelihood function for the Log-logistic distribution}\label{truncated-likelihood-function-for-the-log-logistic-distribution}}
\addcontentsline{toc}{subsubsection}{Truncated likelihood function for the Log-logistic distribution}

The Log-logistic hazard function given log-shape \(\theta_s\), log-scale \(\theta_\eta\) and survival time \(x\) is
\begin{equation*}
h(x, \theta) =  \frac{\frac{e^{\theta_s}}{e^{\theta_\eta}} \left( \frac{x}{e^\theta_\eta} \right)^{e^{\theta_s}-1}}{1 + \left( \frac{x}{e^{\theta_\eta}} \right)^{e^{\theta_s}}},
\end{equation*}

the survivor function is
\begin{equation*}
S(x, \theta) = \frac{1}{1+  \left( \frac{x}{e^{\theta_\eta}}\right)^{e^{\theta_s}}  },
\end{equation*}

and the left-truncated likelihood function is
\begin{align*}
\text{Left-truncated likelihood } p(\mathbf{\tilde  x}|\theta_{s}, \theta_{\eta})
&=\prod_{i=1}^{n_2} \left[ \frac{
\left\{ \frac{\frac{e^{\theta_s}}{e^{\theta_\eta}} \left( \frac{\tilde x_i}{e^\theta_\eta} \right)^{e^{\theta_s}-1}}{1 + \left( \frac{\tilde x_i}{e^{\theta_\eta}} \right)^{e^{\theta_s}}} \right\}^{\tilde\delta_i}
\frac{1}{1+  \left( \frac{\tilde x_i}{e^{\theta_\eta}}\right)^{e^{\theta_s}}  }}{
\frac{1}{1+  \left( \frac{t_1}{e^{\theta_\eta}}\right)^{e^{\theta_s}}}}
\right].
\label{eq:likhood-llogis-trunc}
\end{align*}

\hypertarget{method-for-sampling-from-a-truncated-distribution}{%
\subsubsection*{Method for sampling from a truncated distribution}\label{method-for-sampling-from-a-truncated-distribution}}
\addcontentsline{toc}{subsubsection}{Method for sampling from a truncated distribution}

We can sample values from a truncated survival distribution that lie in the interval \((t_1, \infty)\) as follows. We denote the cumulative density function evaluated at time \(t\) with parameters \(\bm\theta\) as \(F(t, \bm\theta)\).
We first compute the value of the cumulative density function at \(t_1\), \(p=F\left(t_1, \bm\theta\right)\), (i.e.~the probability that a survival time will exceed \(t_1\)). We then sample \(n\) values from a uniform distribution on the interval \([p, 1]\), and plug these into the corresponding \textit{inverse} cumulative density function \(F^{-1}\left(\cdot, \bm\theta\right)\).
This results in \(n\) survival times greater than \(t_1\) that follow the required truncated survival distribution.

\newpage
\beginappendixF

\hypertarget{appendix-f---the-impact-of-increasing-follow-up-durations-on-the-standard-errors-of-the-mcmc-and-gam-estimators}{%
\subsection*{Appendix F - The impact of increasing follow-up durations on the standard errors of the MCMC and GAM estimators}\label{appendix-f---the-impact-of-increasing-follow-up-durations-on-the-standard-errors-of-the-mcmc-and-gam-estimators}}
\addcontentsline{toc}{subsection}{Appendix F - The impact of increasing follow-up durations on the standard errors of the MCMC and GAM estimators}

Increasing follow-up durations affect the standard errors of the nested Monte Carlo and GAM estimators in different ways.
A longer duration of additional follow-up time will result in a greater effective sample size (ESS) of the generated data \(\mathbf{\tilde{x}}\), as the number of observed events \(e\) and time at risk \(y\) will be greater.
When \(\mathrm{ESS} \to 0\), the posterior expectation \(\E_{\bm\theta|\mathbf{x},\mathbf{\tilde{x}}^{(k)}}\) will be similar to the prior expectation \(\E_{\bm\theta}\) for all \(k\), and the variance of the posterior mean will therefore tend to 0.
When \(\mathrm{ESS} \to \infty\), the posterior expectation \(\E_{\bm\theta|\mathbf{x},\mathbf{\tilde{x}}^{(k)}}\) will be similar to the prior parameter sample \(\bm\theta^{(k)}\) that was used to generate the data \(\mathbf{\tilde{x}}^{(k)}\) for all \(k\), and the variance of the posterior mean will therefore converge to the variance of \(\bm\theta\).
Thus, as the variance of the posterior mean increases with increasing values for the additional follow-up time, the standard error of the nested Monte Carlo estimator is expected to increase as well.
The relation between the posterior and prior variance as a function of sample size is further explained in a paper by Jalal \& Alarid-Escudero (2018).\textsuperscript{\protect\hyperlink{ref-jalalGaussianApproximationApproach2018}{40}}

The ESS affects the standard error of the GAM estimator differently.
We recall that the GAM approach relies on expressing the posterior expected net benefit as a function of the generated data \(\mathbf{\tilde{x}}\).
When \(\mathrm{ESS} \to \infty\), the variance of the error term \(\varepsilon^{(k)}\) in the expression \(\NB(d,\bm\theta^{(k)}) = \E_{\bm\theta|\mathbf{x},\mathbf{\tilde{x}}^{(k)}} \{ \NB(d,\bm\theta) \} + \varepsilon^{(k)}\) will tend to 0, since the posterior expectation \(\E_{\bm\theta|\mathbf{x},\mathbf{\tilde{x}}^{(k)}}\) will be similar to the prior parameter sample \(\bm\theta^{(k)}\) that was used to generate the data \(\mathbf{\tilde{x}}^{(k)}\) for all \(k\).
The smaller the variance of the error term \(\varepsilon^{(k)}\), the greater the precision with which the GAM regression coefficients can be estimated, and the smaller the variance of the regression fitted values.
Increasing the length of additional follow-up time increases the precision with which the GAM regression coefficients are estimated, and consequently reduces the standard error of the GAM estimator.

\newpage
\beginappendixG

\hypertarget{appendix-g---maximum-likelihood-estimates-for-the-model-parameters}{%
\subsection*{Appendix G - Maximum likelihood estimates for the model parameters}\label{appendix-g---maximum-likelihood-estimates-for-the-model-parameters}}
\addcontentsline{toc}{subsection}{Appendix G - Maximum likelihood estimates for the model parameters}

\begin{table}
\begin{threeparttable}

\centering

\caption{\label{tab:mle-weib2}Bivariate Normal distribution hyperparameters for the Weibull model parameters given data collected up to \(t_1 = 12\) months}

\begin{tabular}[t]{llrr}
\toprule
\multicolumn{2}{l}{Parameter} & Mean, \(\bm\mu\) & Covariance matrix, \(\bm\Sigma\) \\
\cmidrule(l{3pt}r{3pt}){1-2} \cmidrule(l{3pt}r{3pt}){3-3} \cmidrule(l{3pt}r{3pt}){4-4}
\addlinespace[0.3em]
\multicolumn{4}{l}{\textit{Case study 1: Increasing hazard dataset}}\\
\hspace{1em}
    \(\begin{matrix*}[l]\text{Log shape for new treatment}\\\text{Log scale for new treatment}\end{matrix*}\) & 
    \(\begin{pmatrix*}[r]\theta_{k1}\\\theta_{\lambda 1}\end{pmatrix*}\) & 
    \(\begin{pmatrix*}[r]0.275 \\ 4.014 \end{pmatrix*}\) &
    \(\begin{pmatrix*}[r]0.039 & -0.060 \\-0.060 & 0.117 \end{pmatrix*}\)\\
\addlinespace[0.7em]
\hspace{1em}
    \(\begin{matrix*}[l]\text{Log shape for standard care}\\\text{Log scale for standard care}\end{matrix*}\) & 
    \(\begin{pmatrix*}[r]\theta_{k2}\\\theta_{\lambda 2}\end{pmatrix*}\) & 
    \(\begin{pmatrix*}[r]0.257 \\ 3.863 \end{pmatrix*}\) &
    \(\begin{pmatrix*}[r]0.031 & -0.044 \\-0.044 & 0.081 \end{pmatrix*}\)\\
\addlinespace[0.3em]
\multicolumn{4}{l}{\textit{Case study 2: Decreasing hazard dataset}}\\
\hspace{1em}
    \(\begin{matrix*}[l]\text{Log shape for new treatment}\\\text{Log scale for new treatment}\end{matrix*}\) & 
    \(\begin{pmatrix*}[r]\theta_{k1}\\\theta_{\lambda 1}\end{pmatrix*}\) & 
    \(\begin{pmatrix*}[r]-0.392 \\ 4.472 \end{pmatrix*}\) &
    \(\begin{pmatrix*}[r]0.020 & -0.043 \\-0.043 & 0.136 \end{pmatrix*}\)\\
\addlinespace[0.7em]
\hspace{1em}
    \(\begin{matrix*}[l]\text{Log shape for standard care}\\\text{Log scale for standard care}\end{matrix*}\) & 
    \(\begin{pmatrix*}[r]\theta_{k2}\\\theta_{\lambda 2}\end{pmatrix*}\) & 
    \(\begin{pmatrix*}[r]-0.412 \\4.331\end{pmatrix*}\) &
    \(\begin{pmatrix*}[r]0.018 & -0.036 \\-0.036 & 0.115 \end{pmatrix*}\)\\
\bottomrule
\end{tabular}

\end{threeparttable}
\end{table}

\begin{table}
\begin{threeparttable}

\centering

\caption{\label{tab:mle-gamma}Bivariate Normal distribution hyperparameters for the Gamma model parameters given data collected up to \(t_1 = 12\) months}

\begin{tabular}[t]{llrr}
\toprule
\multicolumn{2}{l}{Parameter} & Mean, \(\bm\mu\) & Covariance matrix, \(\bm\Sigma\) \\
\cmidrule(l{3pt}r{3pt}){1-2} \cmidrule(l{3pt}r{3pt}){3-3} \cmidrule(l{3pt}r{3pt}){4-4}
\addlinespace[0.3em]
\multicolumn{4}{l}{\textit{Case study 1: Increasing hazard dataset}}\\
\hspace{1em}
    \(\begin{matrix*}[l]\text{Log shape for new treatment}\\\text{Log rate for new treatment}\end{matrix*}\) & 
    \(\begin{pmatrix*}[r]\theta_{\alpha1}\\\theta_{\beta 1}\end{pmatrix*}\) & 
    \(\begin{pmatrix*}[r]0.310 \\ -3.752 \end{pmatrix*}\) &
    \(\begin{pmatrix*}[r]0.051 & 0.114 \\0.114 & 0.279 \end{pmatrix*}\)\\
\addlinespace[0.7em]
\hspace{1em}
    \(\begin{matrix*}[l]\text{Log shape for standard care}\\\text{Log rate for standard care}\end{matrix*}\) & 
    \(\begin{pmatrix*}[r]\theta_{\alpha2}\\\theta_{\beta 2}\end{pmatrix*}\) & 
    \(\begin{pmatrix*}[r]0.291 \\ -3.612 \end{pmatrix*}\) &
    \(\begin{pmatrix*}[r]0.042 & 0.088 \\0.088 & 0.208 \end{pmatrix*}\)\\
\addlinespace[0.3em]
\multicolumn{4}{l}{\textit{Case study 2: Decreasing hazard dataset}}\\
\hspace{1em}
    \(\begin{matrix*}[l]\text{Log shape for new treatment}\\\text{Log rate for new treatment}\end{matrix*}\) & 
    \(\begin{pmatrix*}[r]\theta_{\alpha1}\\\theta_{\beta 1}\end{pmatrix*}\) & 
    \(\begin{pmatrix*}[r]-0.434 \\ -4.861 \end{pmatrix*}\) &
    \(\begin{pmatrix*}[r]0.024 & 0.067 \\0.067 & 0.230 \end{pmatrix*}\)\\
\addlinespace[0.7em]
\hspace{1em}
    \(\begin{matrix*}[l]\text{Log shape for standard care}\\\text{Log rate for standard care}\end{matrix*}\) & 
    \(\begin{pmatrix*}[r]\theta_{\alpha2}\\\theta_{\beta 2}\end{pmatrix*}\) & 
    \(\begin{pmatrix*}[r]-0.458 \\-4.752\end{pmatrix*}\) &
    \(\begin{pmatrix*}[r]0.022 & 0.059 \\0.059 & 0.198 \end{pmatrix*}\)\\
\bottomrule
\end{tabular}

\end{threeparttable}
\end{table}

\begin{table}
\begin{threeparttable}

\centering

\caption{\label{tab:mle-lnorm}Bivariate Normal distribution hyperparameters for the Lognormal model parameters given data collected up to \(t_1 = 12\) months}

\begin{tabular}[t]{llrr}
\toprule
\multicolumn{2}{l}{Parameter} & Mean, \(\bm\mu\) & Covariance matrix, \(\bm\Sigma\) \\
\cmidrule(l{3pt}r{3pt}){1-2} \cmidrule(l{3pt}r{3pt}){3-3} \cmidrule(l{3pt}r{3pt}){4-4}
\addlinespace[0.3em]
\multicolumn{4}{l}{\textit{Case study 1: Increasing hazard dataset}}\\
\hspace{1em}
    \(\begin{matrix*}[l]\text{Meanlog for new treatment}\\\text{Log sdlog for new treatment}\end{matrix*}\) & 
    \(\begin{pmatrix*}[r]\theta_{\mu1}\\\theta_{\sigma1}\end{pmatrix*}\) & 
    \(\begin{pmatrix*}[r]4.366 \\ 0.488 \end{pmatrix*}\) &
    \(\begin{pmatrix*}[r]0.164 & 0.062 \\0.062 & 0.029 \end{pmatrix*}\)\\
\addlinespace[0.7em]
\hspace{1em}
    \(\begin{matrix*}[l]\text{Meanlog for standard care}\\\text{Log sdlog for standard care}\end{matrix*}\) & 
    \(\begin{pmatrix*}[r]\theta_{\mu2}\\\theta_{\sigma2}\end{pmatrix*}\) & 
    \(\begin{pmatrix*}[r]4.133 \\ 0.477 \end{pmatrix*}\) &
    \(\begin{pmatrix*}[r]0.113 & 0.044 \\0.044 & 0.023 \end{pmatrix*}\)\\
\addlinespace[0.3em]
\multicolumn{4}{l}{\textit{Case study 2: Decreasing hazard dataset}}\\
\hspace{1em}
    \(\begin{matrix*}[l]\text{Meanlog for new treatment}\\\text{Log sdlog for new treatment}\end{matrix*}\) & 
    \(\begin{pmatrix*}[r]\theta_{\mu1}\\\theta_{\sigma1}\end{pmatrix*}\) & 
    \(\begin{pmatrix*}[r]4.622 \\ 1.047 \end{pmatrix*}\) &
    \(\begin{pmatrix*}[r]0.185 & 0.041 \\0.041 & 0.015 \end{pmatrix*}\)\\
\addlinespace[0.7em]
\hspace{1em}
    \(\begin{matrix*}[l]\text{Meanlog for standard care}\\\text{Log sdlog for standard care}\end{matrix*}\) & 
    \(\begin{pmatrix*}[r]\theta_{\mu2}\\\theta_{\sigma2}\end{pmatrix*}\) & 
    \(\begin{pmatrix*}[r]4.395 \\1.045\end{pmatrix*}\) &
    \(\begin{pmatrix*}[r]0.157 & 0.034 \\0.034 & 0.013 \end{pmatrix*}\)\\
\bottomrule
\end{tabular}

\end{threeparttable}
\end{table}

\begin{table}
\begin{threeparttable}

\centering

\caption{\label{tab:mle-llogis}Bivariate Normal distribution hyperparameters for the Log-logistic model parameters given data collected up to \(t_1 = 12\) months}

\begin{tabular}[t]{llrr}
\toprule
\multicolumn{2}{l}{Parameter} & Mean, \(\bm\mu\) & Covariance matrix, \(\bm\Sigma\) \\
\cmidrule(l{3pt}r{3pt}){1-2} \cmidrule(l{3pt}r{3pt}){3-3} \cmidrule(l{3pt}r{3pt}){4-4}
\addlinespace[0.3em]
\multicolumn{4}{l}{\textit{Case study 1: Increasing hazard dataset}}\\
\hspace{1em}
    \(\begin{matrix*}[l]\text{Log shape for new treatment}\\\text{Log scale for new treatment}\end{matrix*}\) & 
    \(\begin{pmatrix*}[r]\theta_{s1}\\\theta_{\eta 1}\end{pmatrix*}\) & 
    \(\begin{pmatrix*}[r]0.308 \\ 3.915 \end{pmatrix*}\) &
    \(\begin{pmatrix*}[r]0.038 & -0.056 \\-0.056 & 0.107 \end{pmatrix*}\)\\
\addlinespace[0.7em]
\hspace{1em}
    \(\begin{matrix*}[l]\text{Log shape for standard care}\\\text{Log scale for standard care}\end{matrix*}\) & 
    \(\begin{pmatrix*}[r]\theta_{s2}\\\theta_{\eta 2}\end{pmatrix*}\) & 
    \(\begin{pmatrix*}[r]0.297 \\ 3.748 \end{pmatrix*}\) &
    \(\begin{pmatrix*}[r]0.030 & -0.040 \\-0.040 & 0.074 \end{pmatrix*}\)\\
\addlinespace[0.3em]
\multicolumn{4}{l}{\textit{Case study 2: Decreasing hazard dataset}}\\
\hspace{1em}
    \(\begin{matrix*}[l]\text{Log shape for new treatment}\\\text{Log scale for new treatment}\end{matrix*}\) & 
    \(\begin{pmatrix*}[r]\theta_{s1}\\\theta_{\eta 1}\end{pmatrix*}\) & 
    \(\begin{pmatrix*}[r]-0.331 \\ 4.173 \end{pmatrix*}\) &
    \(\begin{pmatrix*}[r]0.019 & -0.037 \\-0.037 & 0.123 \end{pmatrix*}\)\\
\addlinespace[0.7em]
\hspace{1em}
    \(\begin{matrix*}[l]\text{Log shape for standard care}\\\text{Log scale for standard care}\end{matrix*}\) & 
    \(\begin{pmatrix*}[r]\theta_{s2}\\\theta_{\eta 2}\end{pmatrix*}\) & 
    \(\begin{pmatrix*}[r]-0.343 \\4.002\end{pmatrix*}\) &
    \(\begin{pmatrix*}[r]0.017 & -0.030 \\-0.030 & 0.104 \end{pmatrix*}\)\\
\bottomrule
\end{tabular}

\end{threeparttable}
\end{table}

\end{document}